\newif\ifsubmit

\submittrue

\documentclass[twocolumn]{aastex63}

\shorttitle{Mass Budgets and Spatial Scales}
\shortauthors{Mulders et al.}

\usepackage{natbib,graphicx,xspace,amsmath,footnote}
\usepackage{afterpage}

\newcommand{\figp}[1]{(Fig. \ref{f:#1})}

\newcommand{\kepler}{\textit{Kepler}\xspace}

\newcommand{\mearth}[1]{\ensuremath{#1\, M_\oplus}\xspace}
\newcommand{\au}[1]{\ensuremath{#1\, \text{au}}\xspace}

\usepackage{xcolor, fontawesome, microtype}
\definecolor{twitterblue}{RGB}{64,153,255}
\definecolor{linkcolor}{rgb}{0.1216,0.4667,0.7059}

\newcommand{\twitter}[1]{\href{https://twitter.com/#1}{\textcolor{twitterblue}{\faTwitter}\,\tt \textcolor{twitterblue}{@#1}}}

\definecolor{brightmaroon}{rgb}{0.76, 0.13, 0.28}

\definecolor{offwhite}{HTML}{fafaff} 
\pagecolor{offwhite}

\begin{document}

\title{The Mass Budgets and Spatial Scales of Exoplanet Systems and Protoplanetary Disks}

\correspondingauthor{Gijs D. Mulders}
\email{gijs.mulders@uai.cl}


\author[0000-0002-1078-9493]{Gijs D. Mulders}
\affil{Facultad de Ingenier\'ia y Ciencias, Universidad Adolfo Ib\'a\~nez, Av.\ Diagonal las Torres 2640, Pe\~nalol\'en, Santiago, Chile \twitter{GijsMulders}}
\affil{Millennium Institute for Astrophysics, Chile}
\affil{Alien Earths Team, NASA Nexus for Exoplanet System Science, USA} 

\author[0000-0001-7962-1683]{Ilaria Pascucci}
\affil{Lunar and Planetary Laboratory, The University of Arizona, Tucson, AZ 85721, USA}
\affil{Alien Earths Team, NASA Nexus for Exoplanet System Science, USA} 
\author[0000-0002-0093-065X]{Fred J. Ciesla}
\affil{Department of the Geophysical Sciences, The University of Chicago, 5734 South Ellis Avenue, Chicago, IL 60637}
\affil{Alien Earths Team, NASA Nexus for Exoplanet System Science, USA} 

\author[0000-0002-3853-7327]{Rachel B. Fernandes}
\affil{Lunar and Planetary Laboratory, The University of Arizona, Tucson, AZ 85721, USA}
\affil{Alien Earths Team, NASA Nexus for Exoplanet System Science, USA} 

\begin{abstract}
Planets are born from disks of gas and dust, and observations of protoplanetary disks are used to constrain the initial conditions of planet formation. However, dust mass measurements of Class II disks with ALMA have called into question whether they contain enough solids to build the exoplanets that have been detected to date. In this paper, we calculate the mass and spatial scale of solid material around Sun-like stars probed by transit and radial velocity exoplanet surveys, 
and compare those to the observed dust masses and sizes of Class II disks in the same stellar mass regime.
We show that the apparent mass discrepancy disappears when accounting for observational selection and detection biases.
We find a discrepancy only when the planet formation efficiency is below 100\%, or if there is a population of undetected exoplanets that significantly contributes to the mass in solids.
We identify a positive correlation between the masses of planetary systems and their respective orbital periods, which is consistent with the trend between  the masses and the outer radii of Class II dust disks.
This implies that, despite a factor 100 difference in spatial scale, the properties of protoplanetary disks seem to be imprinted on the exoplanet population.
\end{abstract}


\section{Introduction}
The discovery of exoplanets has provided us with an opportunity to better understand the processes by which planets form. In particular, large exoplanet surveys allow to statistically evaluate how common different types of planets and planetary systems are. Transit and radial velocity surveys have shown that exoplanets that are smaller than Neptune are common at  orbital periods less than a year around Sun-like stars \citep{2010Sci...330..653H,2011arXiv1109.2497M,2012ApJS..201...15H,2013ApJ...766...81F}.
A smaller fraction of these sun-like stars, about $10\%$, have giant planets ($> 0.3 M_J$) that are typically located at a few au \citep[e.g.][]{2008PASP..120..531C,2011arXiv1109.2497M,2019ApJ...874...81F,2021ApJS..255...14F}. 

The presence of these exoplanets is challenging to explain based on the measured properties of protoplanetary disks, which are thought to be the environments where planets form.
Surveys of nearby star forming regions at millimeter wavelengths have measured disk dust masses for large numbers of young stars \citep[e.g.][]{2005ApJ...631.1134A}, 
revealing that the typical Class II protoplanetary disk around a Sun-like star has approximately $\sim 10 M_\oplus$ of detectable solids, though with a large dispersion of 0.8 dex \citep[e.g.][]{2016ApJ...828...46A,2016ApJ...831..125P}.
This amount of solids is barely enough to build the exoplanets found in the \kepler and radial velocity surveys \citep{2014MNRAS.445.3315N}.  This implies either that planet formation is {close to 100\%} efficient, or that formation must start in an earlier phase when disks are more massive \citep[e.g.][]{2010MNRAS.407.1981G,2011MNRAS.412L..88G}. 
Indeed, observations of Class I objects indicate a larger dust mass reservoir than Class II disks \citep{2020ApJ...890..130T,2020A&A...640A..19T}. 

\cite{2014MNRAS.445.3315N} carried out the first study comparing the mass budgets of protoplanetary disks and exoplanets, based on the different exoplanet surveys (including \kepler) and disks masses from the Taurus star forming region measured with the Submillimeter Array, SMA \citep{2013ApJ...771..129A}. 
Significant advances have been made since in measuring and characterizing the properties of protoplanetary disks and the demographics of exoplanets. 

For protoplanetary disks, a larger sample of Class II disks have been observed with the Atacama Large Millimeter Array (ALMA) at moderate spatial resolution, extending the census of Class II disks to lower disk and stellar masses \citep[e.g.][]{2016ApJ...827..142B,2016ApJ...828...46A,2016ApJ...831..125P,2017AJ....153..240A,2019MNRAS.482..698C}. 
Surveys have measured disk sizes for a large number of disks, revealing a large population of compact ($< 30$ au) dust disks \citep[e.g.][]{2019ApJ...882...49L,2021A&A...649A..19S}. 
The dust mass and size appear to be tightly correlated, with massive disks being systematically larger than lightweight disks \citep{2017ApJ...845...44T,2018ApJ...865..157A,2020ApJ...895..126H}.

\begin{figure}
    \centering
    \ifsubmit
    \includegraphics[width=0.9\linewidth]{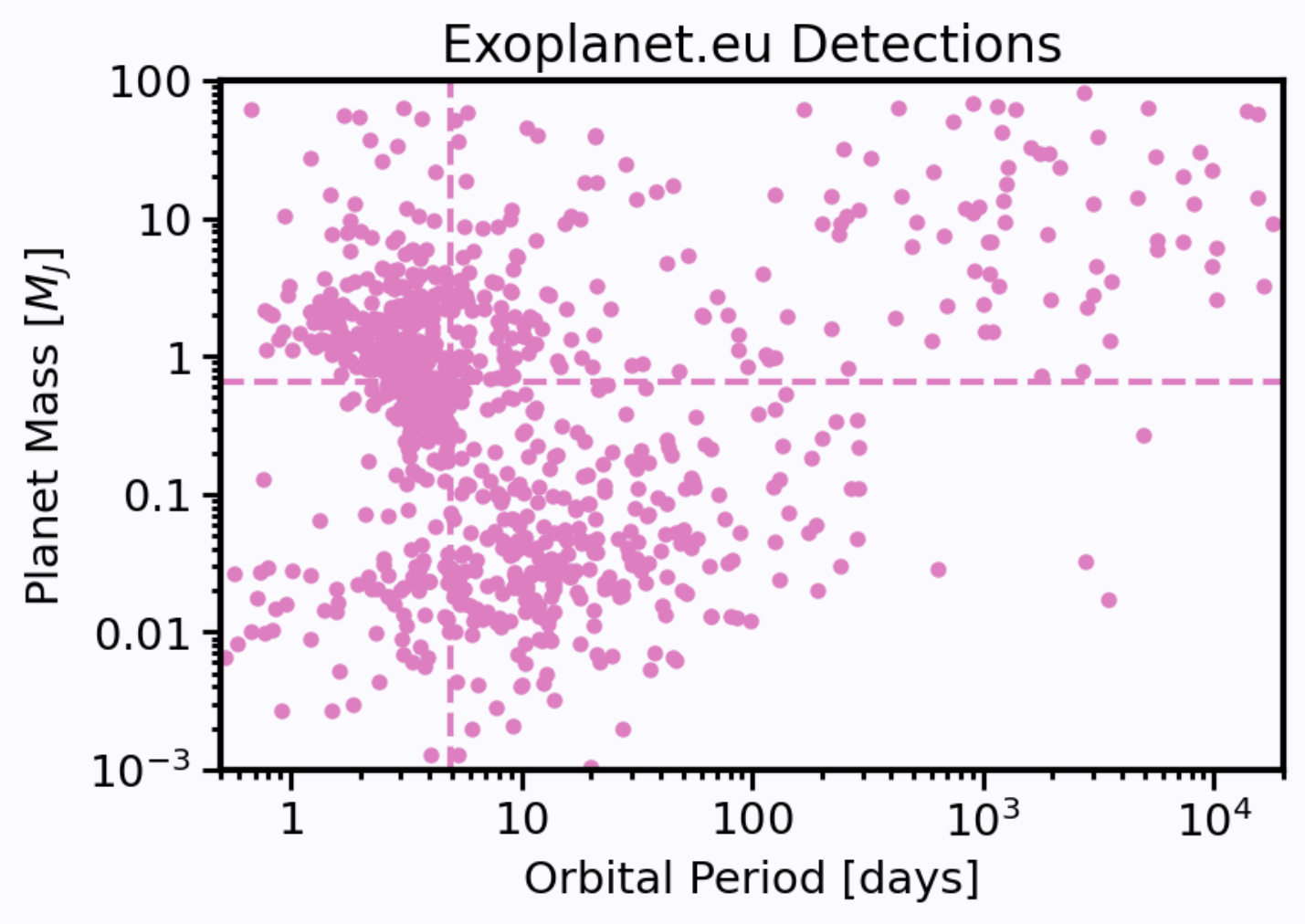} \put(-230,145){a)}\\
    \includegraphics[width=0.9\linewidth]{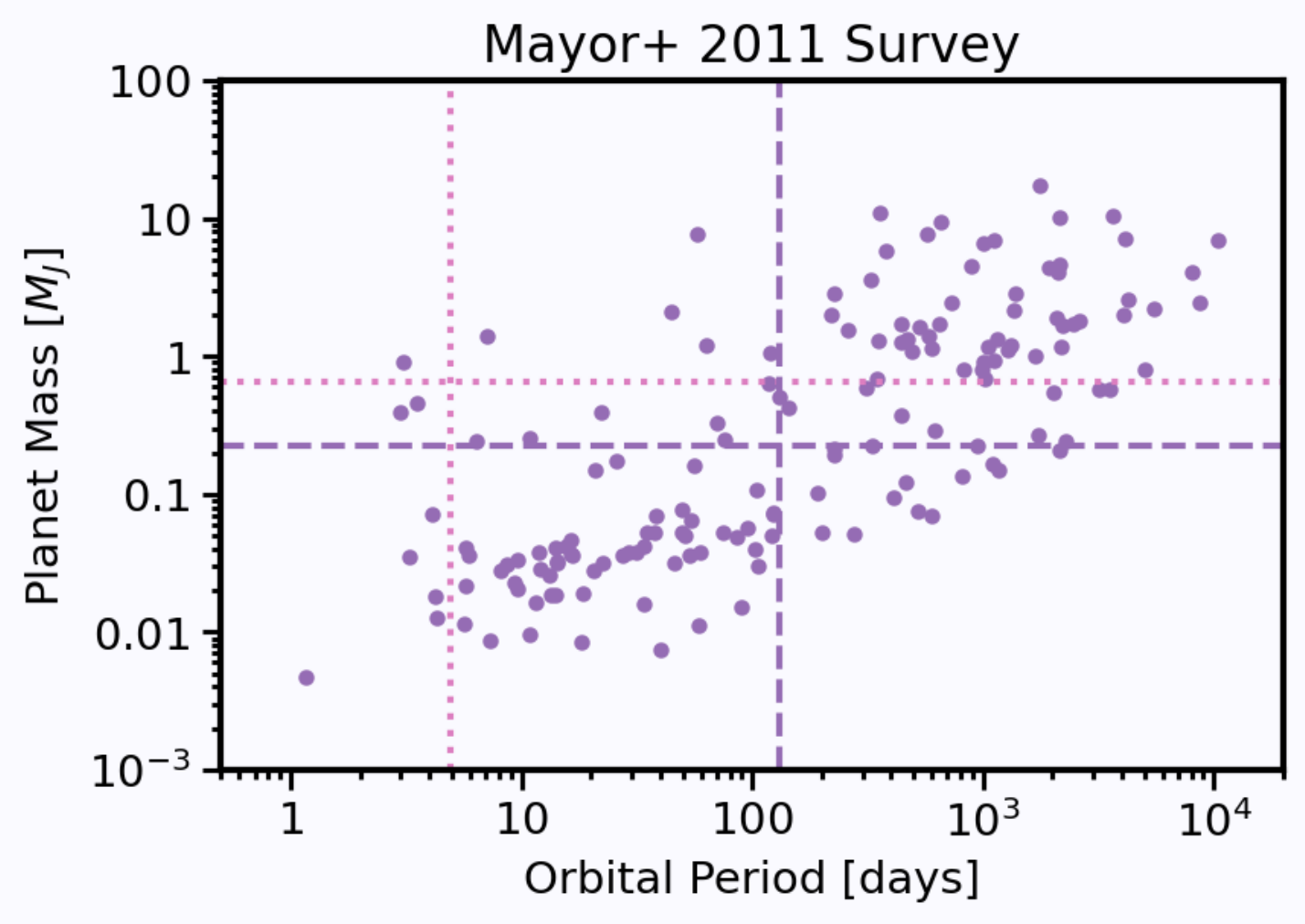} \put(-230,145){b)}\\
    \includegraphics[width=0.9\linewidth]{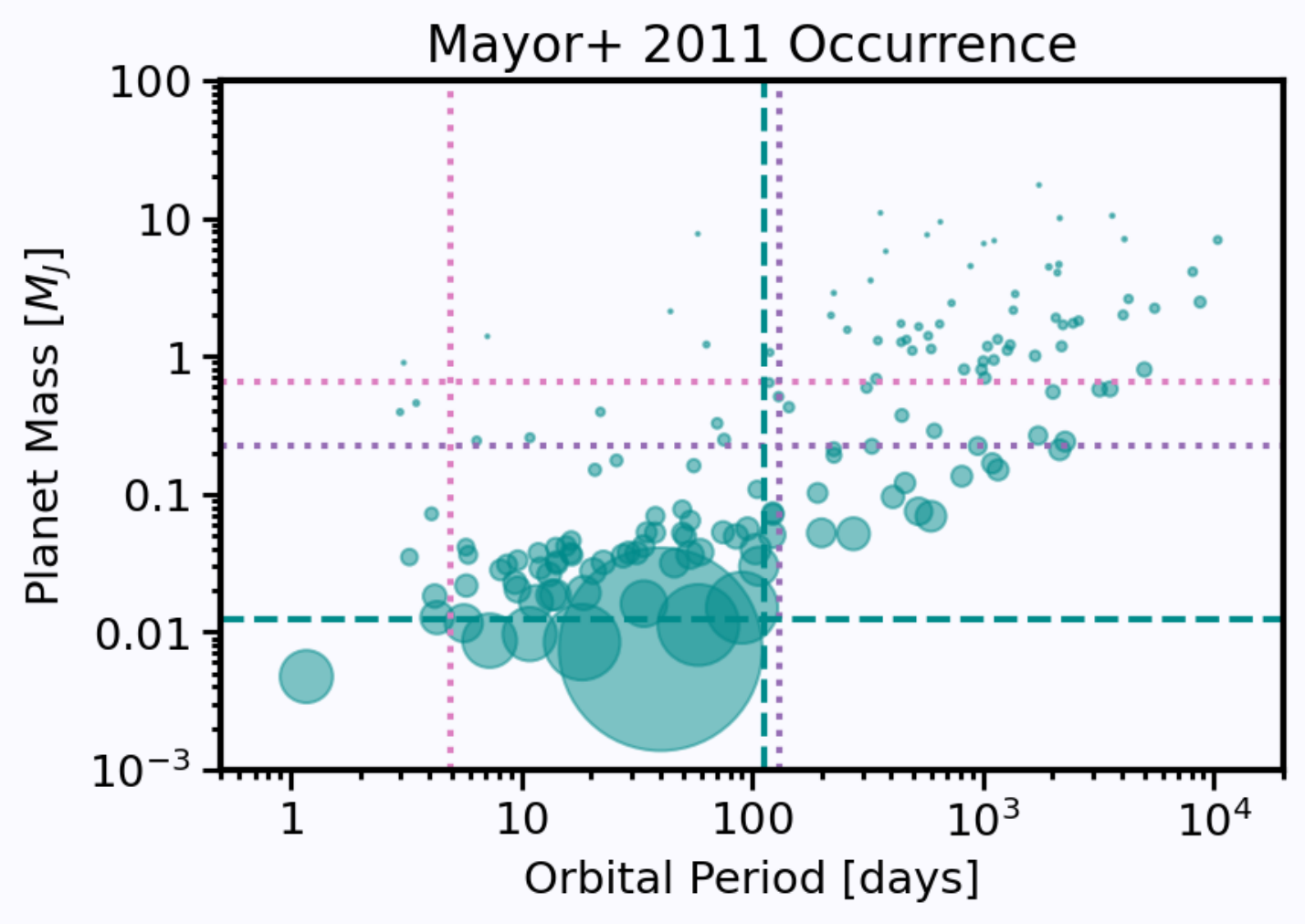} \put(-230,145){c)}\\ 
    \else
    \includegraphics[width=0.9\linewidth]{fig_pdf/fig1a.pdf} \put(-230,145){a)}\\
    \includegraphics[width=0.9\linewidth]{fig_pdf/fig1b.pdf} \put(-230,145){b)}\\
    \includegraphics[width=0.9\linewidth]{fig_pdf/fig1c.pdf} \put(-230,145){c)}\\ 
    \fi
    \caption{
    Exoplanet populations as observed or with bias corrections, as function of planet mass and orbital period. 
    a) Detected exoplanets with a mass measurement from the \url{exoplanet.eu} database, taken on 3/26/2020. 
    b) Detected exoplanets in the volume-limited radial velocity survey from \cite{2011arXiv1109.2497M}.
    c) Same as (b) but correcting for detection bias. The symbol size is scaled proportional to the intrinsic planet occurrence, estimated from the inverse of the detection efficiency as described in \citep{2019ApJ...874...81F}.
    The dotted and dashed lines represent the median planet mass and orbital periods of each dataset.
    The bias correction shows that giant planets, and in particular hot Jupiters, are less frequent than smaller exoplanets.
    }
    \label{f:median}
\end{figure}

On the exoplanet front, the \kepler planet occurrence rates have become more precise after the release of the \texttt{DR25} catalogue \citep{2018ApJS..235...38T} and improved stellar characterization \citep[e.g.][]{2018ApJ...866...99B,Berger:2020jl}. 
With a better understanding of planetary system architectures it is now possible to estimate the fraction of stars with planets as small as the Earth, and statistical reconstructions of planetary system architectures show that $30-50\%$ of Sun-like stars have one or more planets within $1$ au of the star \citep{2018AJ....156...24M,2018ApJ...860..101Z,2019MNRAS.490.4575H}.

At larger distances from the star, radial velocity surveys have measured giant planet occurrence rates out to 10 au, with an estimated occurrence of $10-25\%$ dependent on the lower mass limits (0.1 or 0.3 $M_\text{Jup}$, \citealt{2008PASP..120..531C,2019ApJ...874...81F,2021ApJS..255...14F}. 
Direct imaging surveys provide limits farther out (near $10-100$ au) on the occurrence of massive ($2-13 M_\text{Jup}$) giant planets that are of order $3-6\%$ for solar-mass stars \citep{2019AJ....158...13N,2021A&A...651A..72V}. 
The shape of the semi-major axis distribution, with a peak interior to 10 au measure by radial velocity \citep{2019ApJ...874...81F,2021ApJS..255...14F}
and a declining frequency at larger separations measure by direct imaging \citep{2019AJ....158...13N,2021A&A...651A..72V,2019ApJ...877...46W} indicate that radial velocity surveys may already probe most of the giant planet population.  

\begin{figure*}
    \centering
    \ifsubmit
    \includegraphics[width=0.9\linewidth]{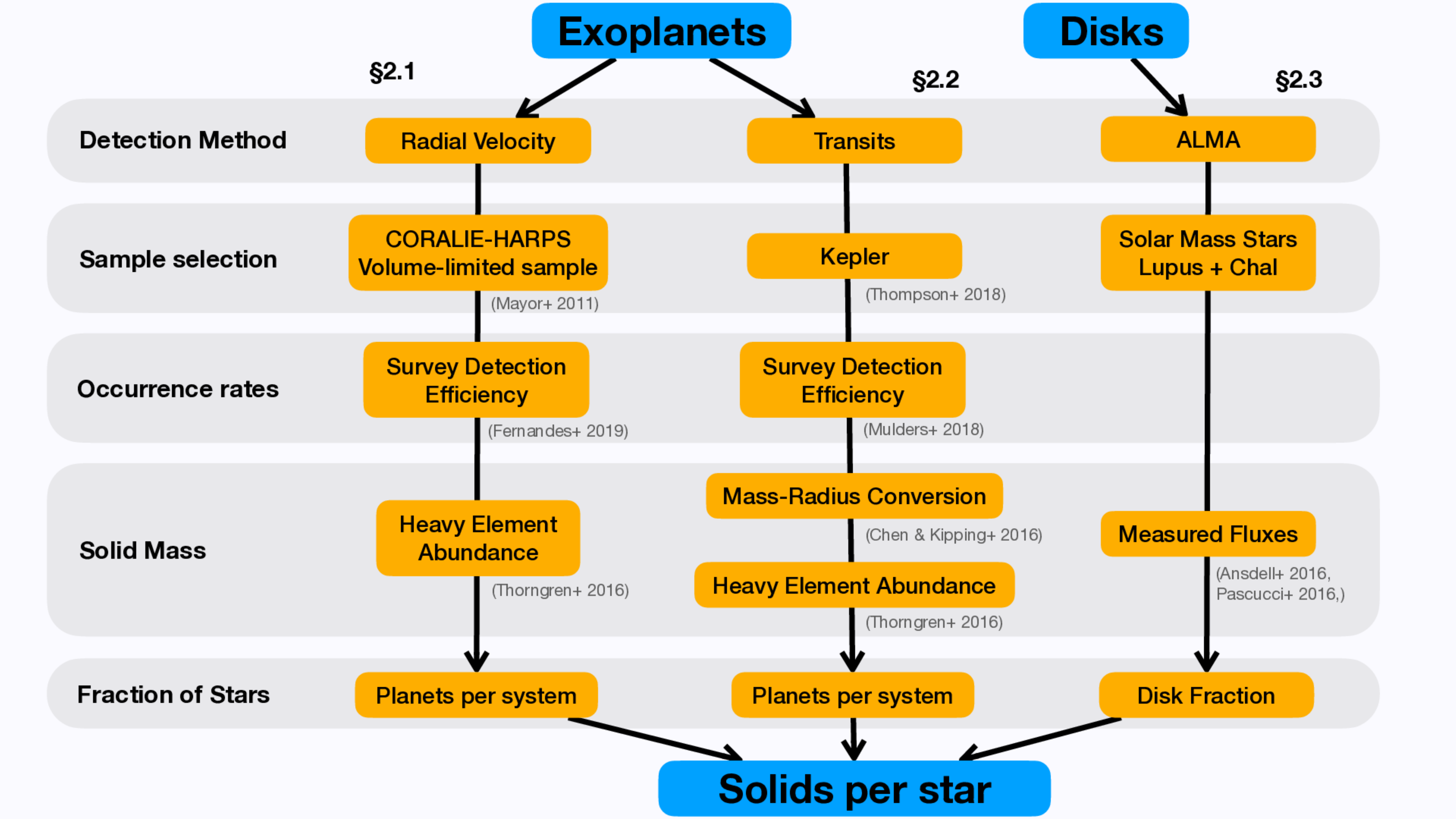}
    \else
    \includegraphics[width=0.9\linewidth]{fig_pdf/fig2.pdf}
    \fi
    \caption{
	Steps taken to de-bias observed samples of exoplanets and disks to facilitate an apples-to-apples comparison between the amount of solids per star. 
    }
    \label{f:sketch}
\end{figure*}

For planets with masses below 0.1 or 0.3 $M_\text{Jup}$ outside of 1 au, the exoplanet census becomes incomplete \citep[e.g.][]{2021ApJS..255...14F}. Microlensing surveys of low-mass stars hint at a large population of Neptune-mass planets between 1-10 au whose occurrence rate is of order unity \citep[e.g.][]{2012Natur.481..167C, 2016ApJ...833..145S}. A direct measurement of the occurrence rate of Earth and Neptune-mass planets at these separations around sun-like stars, however, is not expected to arrive until the \textit{Roman} Space Telescope exoplanet survey \citep[e.g.][]{2019ApJS..241....3P}.

It is now time to re-assess the comparison between disk masses and exoplanets with these new data and insights. 
Recently, \cite{2018A&A...618L...3M} used the collection of ALMA dust disk masses to compare with the masses of detected exoplanets from the online archive \url{exoplanet.eu}. 
The authors found that ``exoplanetary systems masses are comparable or higher than the most massive disks''.
However, the online exoplanet databases do not properly represent the exoplanet population because they contain data from surveys with different sensitivities and targeting strategies. 
As a result, these exoplanets are not representative of the intrinsic exoplanet population and hence, one should be cautious when using them for statistical studies.
For example, hot Jupiters are over-represented because they can be more easily detected in many surveys, which skews the average planet mass to higher values (see Fig. \ref{f:median}).
Thus, looking at only the detected planet masses gives the impression that most planets are as massive as Jupiter, while in reality most planets are Neptune-mass or smaller \citep[e.g.][]{2010Sci...330..653H,2011arXiv1109.2497M}.

In this paper, we derive the mass and spatial scale of solids contained in exoplanets and {Class II} protoplanetary disks, taking into account various detection biases. 
In Section \ref{s:solids}, we determine the mass distribution of solids contained in exoplanets and protoplanetary disks around Sun-like stars. 
Then we calculate the spatial scale of the mass reservoirs probed by detected exoplanets, and compare those to the observed sizes of protoplanetary disks in Section \ref{s:scales}.
We conclude by drawing conclusions on how the statistical links between exoplanets and protoplanetary disks can inform planet formation models.

\section{Solid Mass Budget}\label{s:solids}
In this section, we derive the distribution of solid mass detected in exoplanets and as dust in Class II protoplanetary disks.
A schematic of our approach is shown in figure \ref{f:sketch}.
We focus on Sun-like stars so as to not let the comparisons be biased by stellar-mass dependencies of exoplanet populations \citep[e.g.][]{2018arXiv180500023M} and protoplanetary disks \citep[e.g.][]{2013ApJ...771..129A,2016ApJ...831..125P}.
The two exoplanet samples we consider here have a well-characterized detection bias that allows us to reconstruct the underlying exoplanet population: The \kepler survey of transiting planets \citep[e.g.][]{2010Sci...327..977B,2018ApJS..235...38T} and radial velocity surveys, of which we will use the survey from \cite{2011arXiv1109.2497M}. 
The two exoplanet samples partially overlap for large planets at short orbital periods, but are complementary in other regions of parameter space: the \kepler sample is sensitive to lower-mass planets while the RV sample extends to longer orbital periods.
We omit microlensing surveys \citep[e.g.][]{2016ApJ...833..145S} and direct imaging surveys \citep[e.g.][]{2019AJ....158...13N} here because they have mainly detected exoplanets around lower and higher mass stars. However, we will discuss their planet mass constraints in Section \ref{s:discuss}.

The protoplanetary disk samples are based on the ALMA surveys of Lupus \citep{2016ApJ...828...46A} and Chamaeleon I \citep{2016ApJ...831..125P} of Class II disks.
We omit Class 0/I objects from this analysis because their stellar properties are not known, and thus a potential bias in comparing with exoplanets around Sun-like stars is hard to quantify.
We also do not account for dust in Class III or debris disks, as they contain negligible amounts of measured dust mass \citep[e.g.][]{2008ARA&A..46..339W,2017MNRAS.470.3606H,2020MNRAS.tmp.3151L,2021arXiv210405894M}. 
We will come back to this in the discussion in Section \ref{s:discuss}, where we also discuss the possible contributions of debris disk parent bodies.

Because not all stars have (detected) disks or exoplanets, we will calculate the solid mass distribution as a fraction of all stars.

\subsection{Radial Velocity Exoplanets}
\begin{figure}
    \centering
    \ifsubmit
    \includegraphics[width=\linewidth]{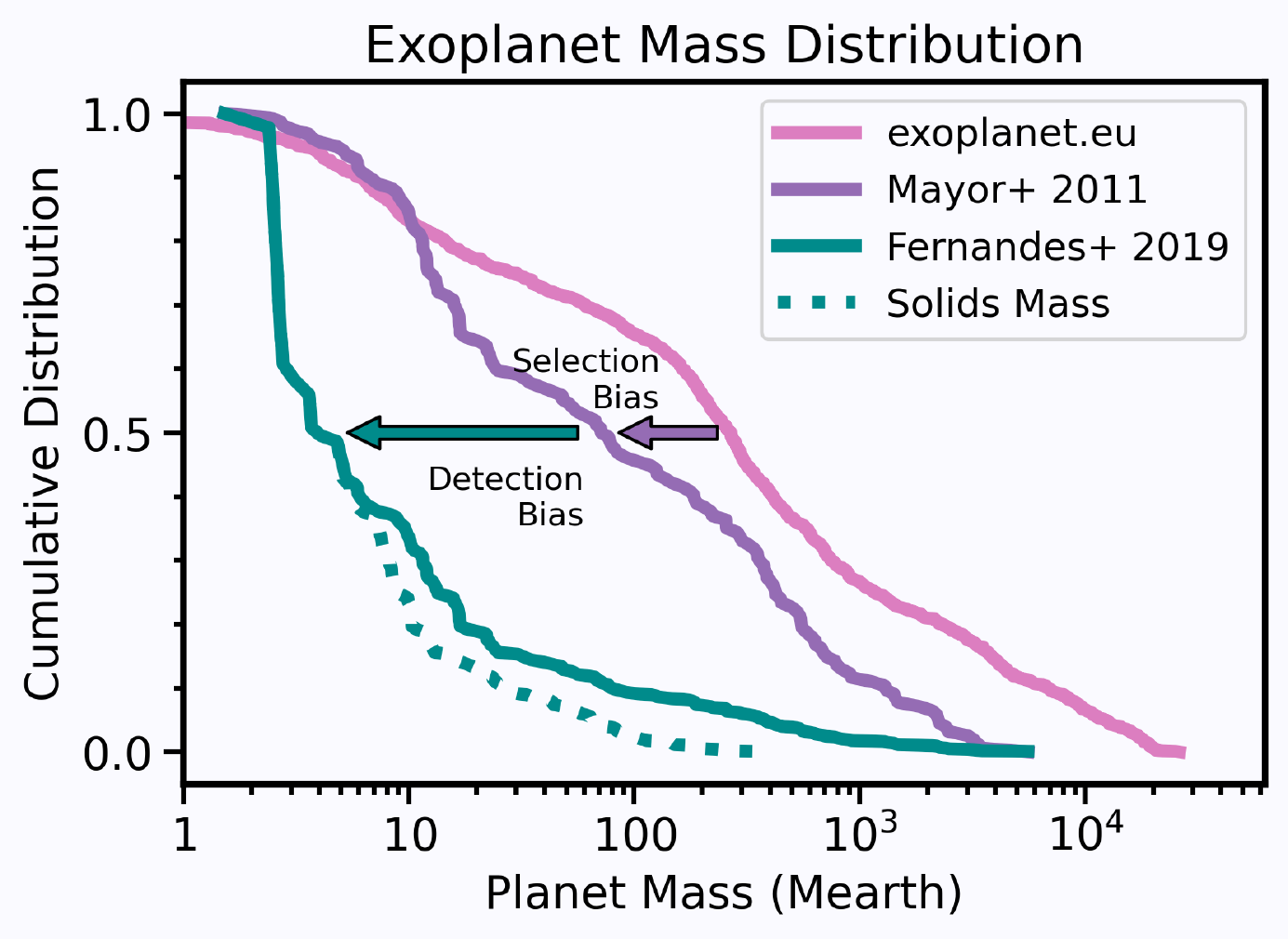}
    \else
    \includegraphics[width=\linewidth]{fig_pdf/fig3.pdf}    
    \fi
    \caption{
    Selection and detection biases affect the exoplanet mass distributions {from Fig. \ref{f:median}}, as indicated by arrows.
    The pink line represents the cumulative mass distribution of all known exoplanets with a reported mass from the exoplanet.eu database, taken on 3/26/2020. 
    The purple line represents the detected mass distribution from the volume-limited radial velocity survey from \cite{2011arXiv1109.2497M}.
    The teal line represents the detection bias-corrected planet mass distribution from \citep{2019ApJ...874...81F}.
    The teal dotted line represents the \textit{solid} mass distribution estimated using the heavy element mass fraction scaling relation from \cite{2016ApJ...831...64T}.
    }
    \label{f:bias}
\end{figure}

We use the CORALIE--HARPS survey of \cite{2011arXiv1109.2497M} to calculate the {solid} mass distribution of radial velocity detected exoplanets. This survey was conducted using a volume-limited sample of stars similar in mass to the Sun. It is sensitive to planets as low in mass as $\sim 3\,M_\oplus$ at short orbital periods, while the largest planets are detected out to a distance of $\sim{}10$ au {(Fig. \ref{f:median}b)}.  
The detected planet mass distribution is shown with the purple line in Figure \ref{f:bias}, with a median detected planet mass of $\sim{100}\,M_\oplus$.

To highlight the effects of the stellar sample selection, the planet mass distribution of all detected exoplanets from the online archive \texttt{exoplanet.eu} {that was used by \cite{2018A&A...618L...3M}} is shown in pink in Fig. \ref{f:bias}. 
This sample is a heterogeneous combination of exoplanets surveyed with different sensitivities in terms of planet mass. 
{Hot Jupiter are over-represented in this sample which} leads to a median planet mass of $300\,M_\oplus$ that is a factor of 3 higher than the volume-limited stellar sample of \cite{2011arXiv1109.2497M} survey, as indicated by the purple arrow {in Figure \ref{f:bias} and the horizontal lines in Figure \ref{f:median}}. 

Even within a single survey, the sensitivity to detect planets of different masses varies from star to star.
The survey detection efficiency as function of planet mass and orbital period was characterized by \cite{2011arXiv1109.2497M} using injection-recovery tests. For each detected exoplanet, $j$, we extract the survey completeness, $C_j$, as described in \cite{2019ApJ...874...81F}. 
This completeness is defined as the probability that the planet is detected in the survey based on its mass and orbital period. 
We calculate a weight factor, $w_j=\frac{1}{N_\star\,C_j}$, where $N_\star$ is the number of stars in the survey, to de-bias the mass distribution of exoplanets ({Fig. \ref{f:median}c and} Fig. \ref{f:bias}, teal line). The median planet mass is $\sim 5\,M_\oplus$, 
which is more than an order of magnitude lower than the median mass of the biased distribution of detections, as indicated by the teal arrow. Accounting for the combined detection and selection effects yields a median mass that is almost two orders of magnitude lower than the median mass of all known exoplanets {from the online archive \texttt{exoplanet.eu}}.

The more massive planets are gas giants and thus {a significant fraction of their measured mass is not in solids}. To calculate the solid components in these exoplanets, $M_c$, we use the best-fit scaling relation between heavy-element abundance and planet mass from \cite{2016ApJ...831...64T}, given by:
\begin{equation}\label{eq:mcore}
M_c=  
\begin{cases}
60\, M_\oplus \left(\frac{M}{M_J}\right)^{0.6} & \text{if}\, M>M_c \\
M & \text{otherwise}
\end{cases}
\end{equation}
The correction for planet heavy element mass primarily affects the more massive end of the planet mass distribution (Fig. \ref{f:bias}, teal dotted line).
{We ignore the intrinsic dispersion in the relation because we are primarily interested in the core mass distribution, which is less affected by scatter than estimates for individual planets.}

A final correction that has to be made before exoplanets can be compared to disks is to calculate what fraction of stars have planets of a given mass. Some stars may not have planetary systems at all, or some planets may be below the survey detection limit. Assuming the mass distribution of detected exoplanets applies to all stars can significantly bias the results, especially if planet occurrence rates are much smaller than one. For example, giant planets are typically detected around $\sim 10\%$ of stars \citep[e.g.][]{2008PASP..120..531C,2019ApJ...874...81F}, so as long as 10\% of stars have similarly massive disks, the average disk mass be much lower without creating a conundrum.

The fraction of stars with planetary systems, here denoted by $F$, is related to the planet occurrence rate, $f = \sum_j w_j$, by: 
\begin{equation}
F = \frac{f}{\bar{n}}
\end{equation}
where $\bar{n}$ is the average number of planets per planetary system. The number of planets per system has not been robustly determined from radial velocity surveys. The average detected number of planets per system is 1.5 in the sample of \cite{2011arXiv1109.2497M}, but this number increases for the smaller planets that are intrinsically more common, so this number is likely a lower limit.
Therefore, we assume a somewhat higher number of two planets per system ($\bar{n}_\text{RV}=2$).
We note that the detection efficiency of multi-planet systems was not considered in the injection-recovery tests of \cite{2011arXiv1109.2497M}, and thus it is difficult to determine the intrinsic planet multiplicity based on publicly available data.

The mass in solids per planetary system, $M_s$, is estimated from each individual planet detection as $M_{s} = \bar{n}_\text{RV} \, M_{c}$.
The fraction of stars with at least a given amount of solids in its planetary system is then:
\begin{equation} 
F(>M_s)= \frac{1}{\bar{n}_\text{RV}} \sum_{j, M_{sj}>M_s} M_{sj} w_j
\end{equation}
{After applying these corrections, the fraction of stars with planetary systems more massive than $10 M_\oplus$ is $32\%$, e.g. the radial velocity planet population represents roughly a third of solar-mass stars.}.

\begin{figure}
    \centering
    \ifsubmit
    \includegraphics[width=\linewidth]{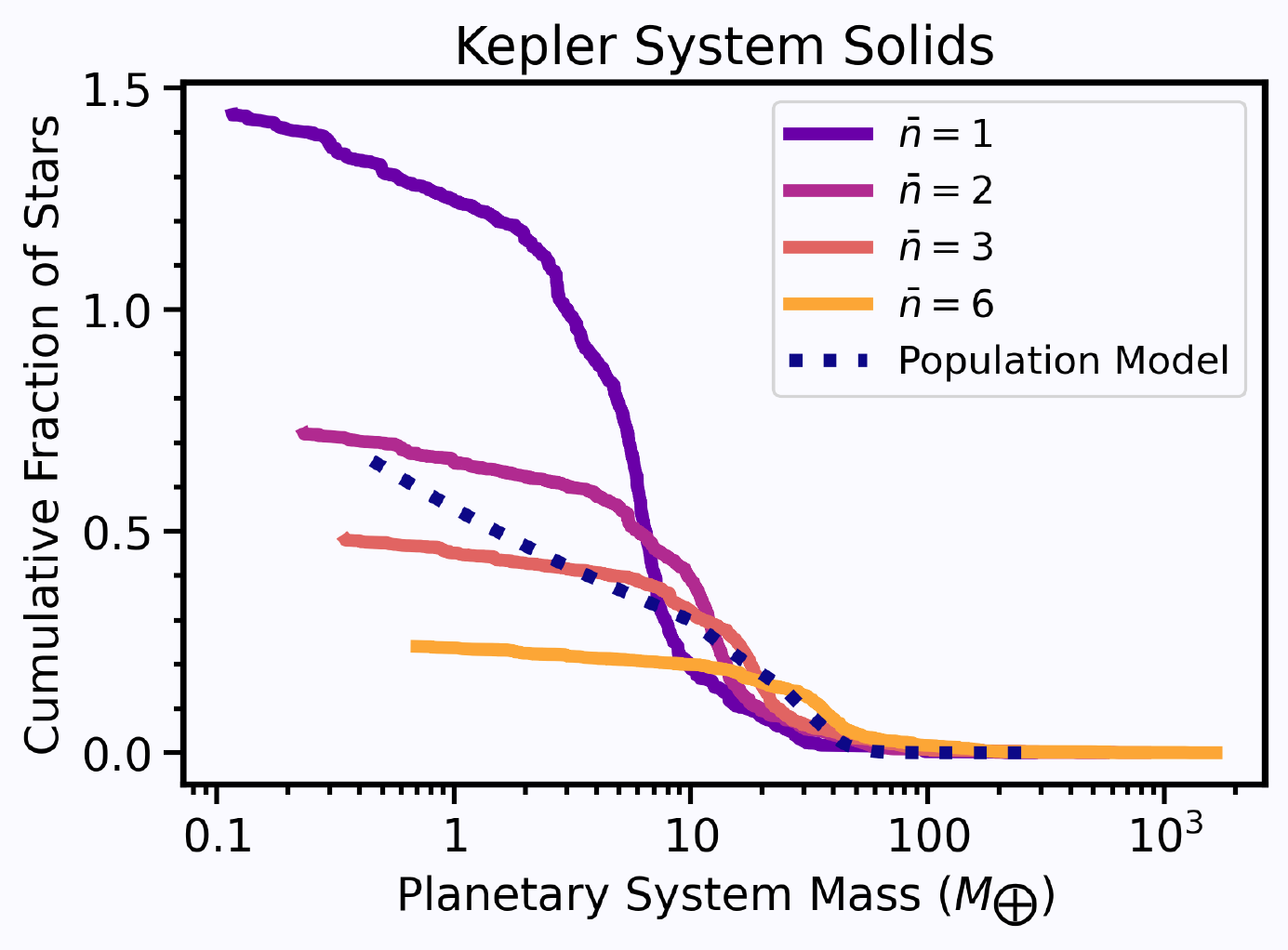}
    \else
    \includegraphics[width=\linewidth]{fig_pdf/fig4.pdf}
    \fi
    \caption{
    Fraction of stars with planets, calculated from the \kepler planet occurrence rate under different assumptions of the number of planets per system, $\bar{n}$.
    The distribution is compared to the parametric population model from \cite{2018AJ....156...24M} which makes an assumption of a broken power-law radius distribution. We adopt $\bar{n}=3$ throughout this paper.
    Note that $\bar{n}=1$ yields a fraction above 1 which is not a physical solution. 
    }
    \label{f:bias_multi}
\end{figure}

\subsection{Transiting Exoplanets from Kepler}
We derive the mass distribution of transiting exoplanets based on the survey completeness from \cite{2018AJ....156...24M}. This completeness was calculated on a grid of orbital period and planet radius using all main-sequence stars in the \kepler survey. This stellar sample has a median mass, estimated from isochrone fitting using \texttt{isoclassify} \citep{2017ApJ...844..102H}, of $1.0$ solar mass and a standard deviation of $0.3\,M_\odot$, and its occurrence rate is representative of that of solar-mass stars. We therefore do not make any additional mass or effective temperature cuts, as they would only reduce the sample size. %
The calculation uses the detected planet candidates from the \kepler \texttt{DR25} planet catalog \citep{2018ApJS..235...38T}, the survey detection efficiency measurements from \texttt{KeplerPORTS} \citep{2017ksci.rept...19B}, and the \textit{Gaia} stellar parameters \citep{2018ApJ...866...99B}. For each detected planet candidate, we determine the weight factor $w_j = \frac{1}{N_\star\,C_j}$ by evaluating the survey completeness at the period and radius of the planet. 

Next, we calculate the planet mass from the detected planet radii using the best-fit mass-radius relation from \cite{2017ApJ...834...17C}:
\begin{equation}\label{eq:mradius}
{
M (M_\oplus) = 
\begin{cases}
\left(\frac{R}{R_\oplus}\right)^{1/0.28} & \text{if}\, R< R_t \\
M_t \left(\frac{R}{R_t}\right)^{1/0.59} & \text{otherwise}
\end{cases}
}
\end{equation}
{with $M_t= 2 M_\oplus$ and $R_t=(M_t/M_\oplus)^{0.28} R_\oplus$.}
{While the mass-radius relation in \cite{2017ApJ...834...17C} is probabilistic, we ignore this aspect since we are considering the wider distribution of planet properties which is less affected by intrinsic scatter.}
The mass in solids per planet is again calculated using the estimates of heavy-element abundance from \cite{2016ApJ...831...64T} in Eq. \ref{eq:mcore}. The planet solid mass distribution is shown in Figure \ref{f:bias_multi} with the solid dark purple line. 
We note that the cumulative occurrence of planets adds up to a number larger than 1, showing that a correction for multiple planets per star is necessary. 

As a final step, we calculate the fraction of stars with the given amount of planetary mass in solids by correcting for the number of planets per system, $n_j$.
The fraction of stars with planets, $F$, is related to the individual planet weights as:
\begin{equation}
F= \sum_j \frac{w_j}{n_j}
\end{equation}
where $n_j$ is the number of planets in the system. Unfortunately, in transit surveys $n_j$ can not be directly measured because of geometrical biases: the probability that multiple planets transit their star is low even for low mutual inclinations of the planets orbits. Thus, the number of detected transiting planets around a star is not a good measure of the real multiplicity because most planets in a system will not be transiting.

However, the \emph{average} number of planets per system, $\bar{n}$, can be estimated using a careful treatment of the complex geometric detection biases \citep[e.g.][]{2011ApJS..197....8L,2012AJ....143...94T}.
The mass in solids per planetary systems is approximated as $M_s = \bar{n} \, M_c$. 
The previous equation can then be simplified as:
\begin{equation}
F= \frac{1}{\bar{n}} \sum_j w_j 
\end{equation}
and the fraction of stars with at least a given amount of solids is:
\begin{equation}
F(>M_s)= \frac{1}{\bar{n}} \sum_{M_{sj}>M_s} w_j
\end{equation} 

Figure \ref{f:bias_multi} shows the fraction of stars with at least a given amount of solids for different values of $\bar{n}= 1,2,3,6$. 
Forward modeling of \kepler planet populations indicates that $\bar{n}$ is typically in the range $2-6$ \citep{2018AJ....156...24M,2018ApJ...860..101Z,2019MNRAS.490.4575H}, though the intrinsic distribution of planet multiplicity is uncertain and model-dependent \citep{2018ApJ...860..101Z,2019MNRAS.489.3162S,2019MNRAS.483.4479Z}. 

To identify the best value of $\bar{n}$ for the \kepler sample, we compare the different distributions to the fraction of stars with planetary systems estimated using a parametric forward model from \cite{2018AJ....156...24M} (dashed line). We find that the assumption $\bar{n}=3$ works well for the \kepler sample. 
This yields a cumulative fraction of Sun-like stars with detectable planets of $F=47\%$ \figp{cdf} which is also consistent with estimates from other forward models \citep[e.g.][]{2018ApJ...860..101Z,2019MNRAS.490.4575H,2020AJ....159..164Y}.

\begin{figure}
    \centering
    \ifsubmit
    \includegraphics[width=\linewidth]{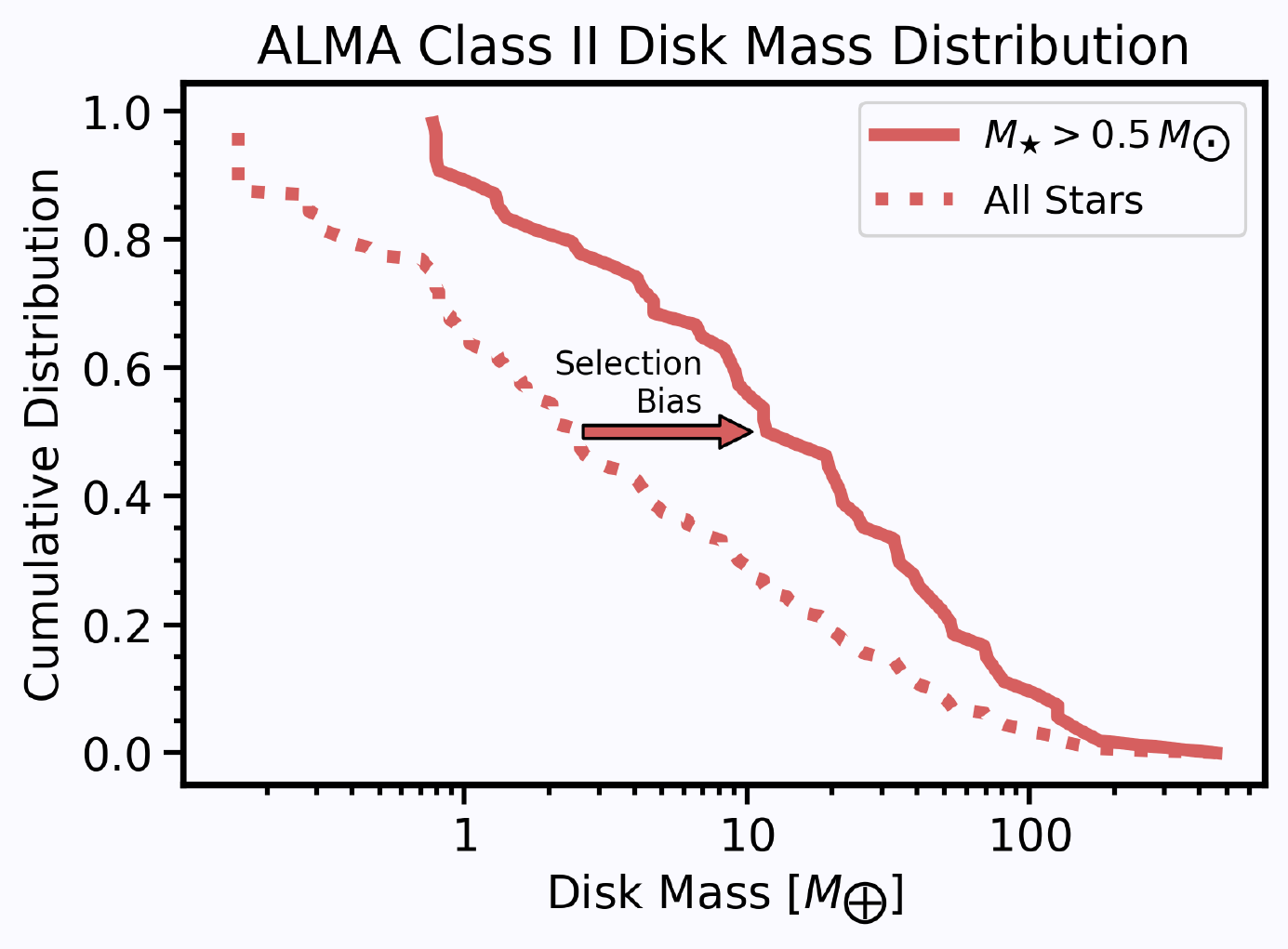}    
    \else
    \includegraphics[width=\linewidth]{fig_pdf/fig5.pdf}    
    \fi
    \caption{
    Stellar selection biases affect the protoplanetary disk mass distributions, indicated by the arrow.
    The dotted red line shows the disk mass distribution of all stars in Lupus and Chamaeleon I from \cite{2017ApJ...847...31M}. The solid line shows how the mass distribution changes when only selecting Sun-like stars, here defined as $0.5 < M_\star/M_\odot < 2$.
    }
    \label{f:bias_disks}
\end{figure}

\subsection{Protoplanetary Disk Dust Masses}
Solid masses of {Class II} protoplanetary disks have been measured with ALMA in multiple star forming regions \citep[e.g.][]{2016ApJ...827..142B,2016ApJ...828...46A,2016ApJ...831..125P,2017AJ....153..240A,2019MNRAS.482..698C}. These mass measurements represent the amount of solids in millimeter sized dust grains, and are calculated from the millimeter flux density using the stellar distance and assumptions on the disk temperature, dust opacity, and optical depth. 
Here we use the combined Chamaeleon I \citep{2016ApJ...831..125P} and Lupus \citep{2016ApJ...828...46A} stellar samples because for these regions stellar masses have also been homogeneously re-computed \citep[e.g.][]{2016ApJ...831..125P,2017A&A...604A.127M,2017A&A...600A..20A}. 
We use the disk masses from the online table of \cite{2017ApJ...847...31M}, calculated using an average disk temperature of $T=20 K$. The cumulative distribution of disk masses is shown with the dotted line in Figure \ref{f:bias_disks}. From this sample, we select only disk with approximately solar-mass host stars ($0.5 < M_\star/M_\odot < 2$). 
This cut mainly removes low-mass stars because they are more numerous than Sun-like stars.
Because of the positive correlation between protoplanetary disk mass and stellar mass, this cut increases the median disk mass by a factor five from $\sim 2 M_\oplus$ to $\sim 10 M_\oplus$.

\begin{figure}
    \centering
    \ifsubmit
    \includegraphics[width=\linewidth]{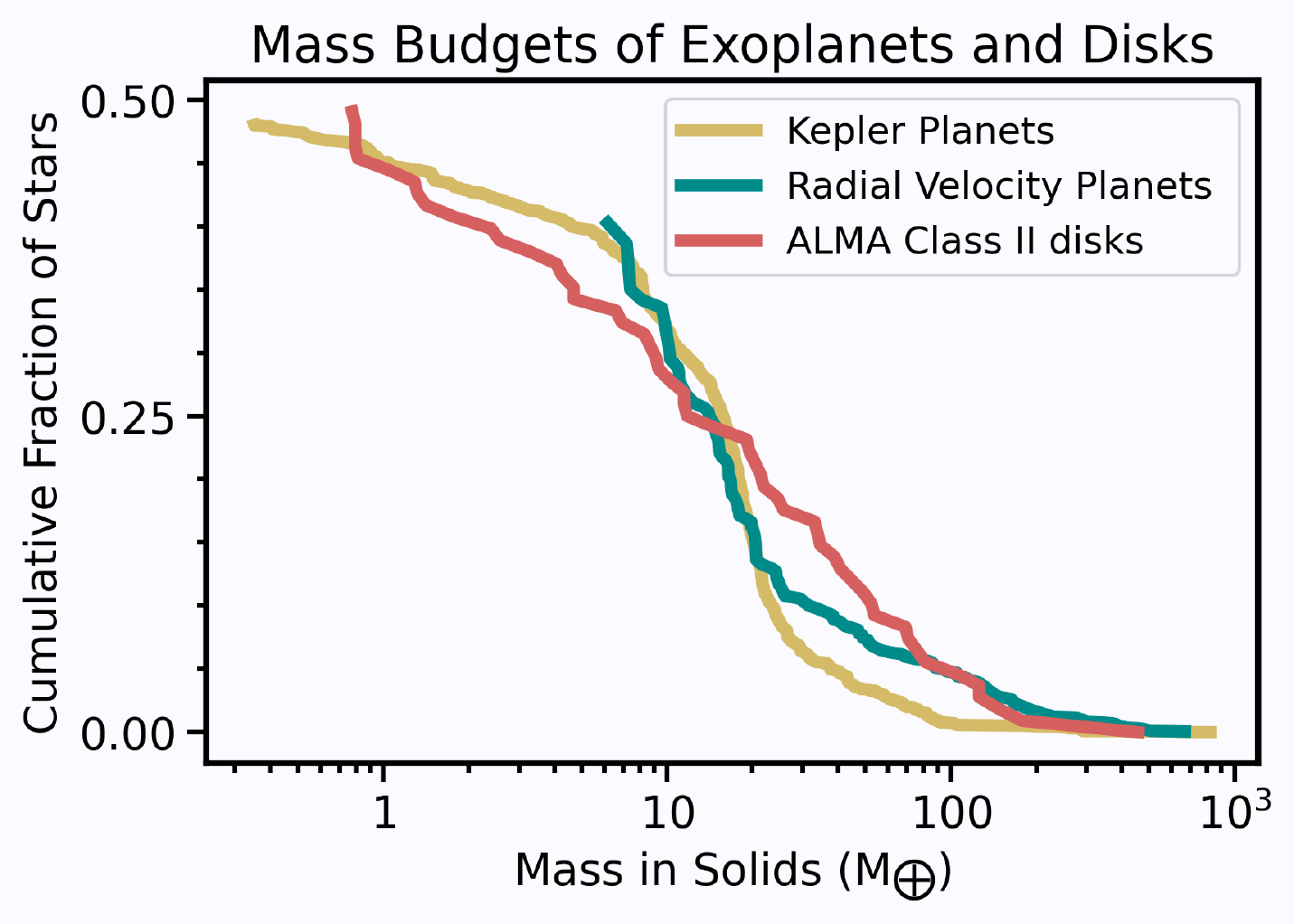}    
    \else
    \includegraphics[width=\linewidth]{fig_pdf/fig6.pdf}    
    \fi
    \caption{
    Solid mass budget of exoplanet systems and protoplanetary disks around solar-mass stars.
    The distributions for the \kepler and radial velocity surveys are corrected for detection biases, while the protoplanetary disk surveys are complete down to very low disk mass.
    The cumulative distributions are expressed as a fraction of all stars. 
    Note that not all stars have disks or exoplanets, at least within the detection limits of current surveys.
    }
    \label{f:cdf}
\end{figure}

Stars without protoplanetary disks (Class III objects) are excluded from these surveys, and a correction has to be made to estimate the fraction of stars with a given amount of solids.
{Here, we assume that Class III disks have a negligible amount of detectable dust compared to Class II objects, as is the case for Lupus at least \citep{2020MNRAS.tmp.3151L}.}
The fraction of stars with protoplanetary disks decreases with time \citep[e.g.][]{2001ApJ...553L.153H} and is typically $50\%$ at the age of Lupus and Chamaeleon I \citep{2005ApJ...631L..69L}.
Therefore, we normalize the mass distribution of disks such that the cumulative number of stars with disks of any mass is $F_\text{disk}= 50\%$ \figp{cdf}. 
{Because the disk fraction tends to decrease with time, the CDF of an older(younger) star forming region would have a lower(higher) normalization.}

\subsection{Comparing Disk and Exoplanet Masses}
Figure \ref{f:cdf} shows the distribution of solid masses derived from samples of transiting planets, radial velocity planets, and {Class II} protoplanetary disks.  

\begin{figure}
    \centering
    \ifsubmit
    \includegraphics[width=\linewidth]{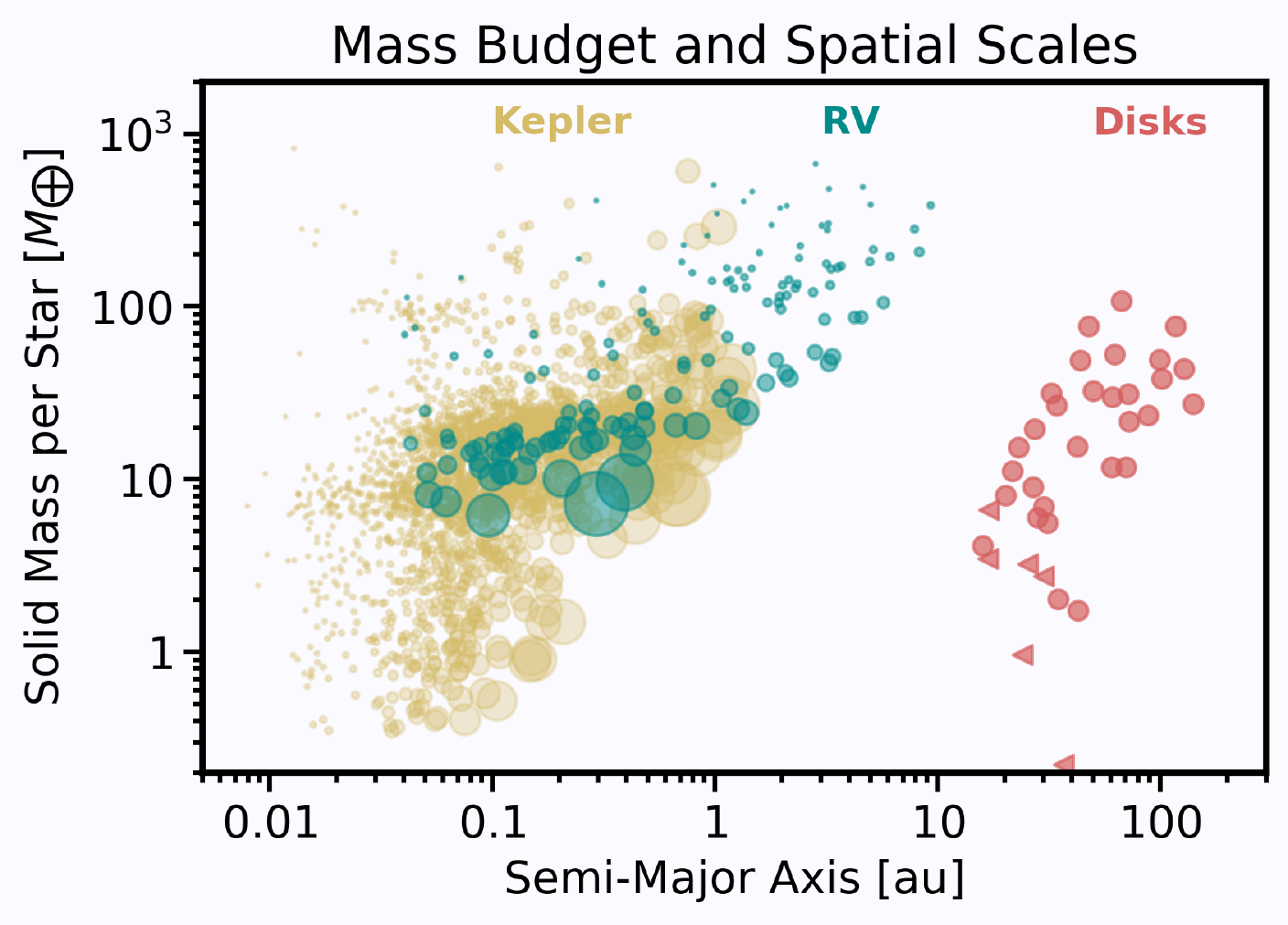}
    \else
    \includegraphics[width=\linewidth]{fig_pdf/fig7.pdf}
    \fi    
    \caption{
    Estimated solid system mass vs. spatial scale for exoplanets systems and protoplanetary disks around solar-mass stars. The symbol size is proportional to the inverse survey completeness for each exoplanet to better reflect the true occurrence.
    The sizes of protoplanetary disks are the radii that enclose 68\% of the flux (circles), or an upper limit to that value (triangle). 
        }
    \label{f:PR}
\end{figure}

The \kepler and radial velocity surveys overlap in the range $5< M_s/M_\oplus < 20$ where both surveys trace the same population of exoplanets in terms of mass and period (see also Fig. \ref{f:PR}). At larger planetary system solid masses, the radial velocity planets trace an additional population of colder giants outside of $1$ au that is not detected with \kepler, and the distributions deviate.

All three distributions peak near $\approx 10\,M_\oplus$, consistent with previous estimates of the average amount of solids in planets and disks \citep[e.g.][]{2014MNRAS.445.3315N,2015ApJ...814..130M,2016ApJ...831..125P}. 
The distribution of disk masses appears wider than that of exoplanets. 
However, we do not see many exoplanetary systems whose masses are higher than the most massive disks as claimed by \cite{2018A&A...618L...3M}, which we attribute to the authors not taking into account necessary bias corrections for exoplanets. {While a number of planetary systems exist with solid masses above $100 M_\oplus$ -- a region where few disks are detected -- these constitute a tiny fraction of the total systems in the tail of the probability distribution (see Fig. \ref{f:cdf}), and do not constitute a significant mass budget issue at the population level.}

We note that the fraction of stars with disks and planets never go above $50\%$, so our analysis is only accounting for half the sun-like stars. The other stars may either have exoplanets outside of the detection limits of transit and radial velocity surveys or never have formed planets at all. The observed fraction of stars with disks depend on the evolutionary phase, and typically decreases with time \citep[e.g.][]{2001ApJ...553L.153H}. It is therefore unlikely that protoplanetary disks in older regions like Upper Sco, with a disk fraction below 25\% \citep{2012ApJ...758...31L}, represent the formation sites of all observed exoplanets. While planet formation is likely ongoing in the disks that survive to this age, more than half the exoplanet systems must have formed around stars that have already dispersed their disks by the age of Upper Sco, 5-10 Myr. 

Younger regions, such as Taurus, with a higher disk fraction of 75\% \citep{2010ApJS..186..111L}, therefore include a significant fraction of disks that do not form the exoplanets that can be detected with radial velocity or transit surveys. Either some of these disks form different types of planets (such as traced by micro-lensing or debris) or some of them may not form exoplanets at all.

In general, there is relatively good agreement between the different solid mass distributions. This result supports the earlier conclusion from \cite{2014MNRAS.445.3315N} that the mass reservoirs of solids contained in the \kepler and radial velocity exoplanets are of similar magnitude as those in protoplanetary disks. 

\section{Spatial Scales}\label{s:scales}
The solid mass reservoirs of exoplanets and protoplanetary disks, while similar in magnitude, are detected at different distances from the star. Dust disk sizes are typically in the range 10-100 au, with most of the mass located at larger radii. On the other hand, most exoplanets are detected at smaller radii, with giant planets mainly concentrated between 1-10 au and smaller planets detected with high occurrence interior to 1 au, though their census at larger radii is incomplete. 
 
 In this section, we will quantify the difference between the measured outer radii of protoplanetary disks and the observed location of the exoplanet population. 
In particular, we will focus on the positive correlation between disk mass and size identified by \cite{2017ApJ...845...44T} and others. Massive Class II protoplanetary disks are observed to be larger than lighter disks, which in turn tend to be more compact. We will look for a similar trend in the exoplanet population by measuring the sizes of planetary systems as a function of their solid mass.
 
We note that this analysis of measuring the spatial scales of solids is different than that of the Minimum Mass Solar Nebula \citep{1977Ap&SS..51..153W,1981PThPS..70...35H} or Minimum Mass Extra-Solar Nebula \citep[e.g.][]{2004ApJ...612.1147K}.
The MMSN or MMEN attempt to reconstruct the mass (or surface density) distribution within protoplanetary disks based on the masses and locations of (exo)planets within planetary systems. This approach has also been applied directly to exoplanet populations \citep[e.g.][]{2013MNRAS.431.3444C}.
In contrast, in the spatial scales analysis of this paper we focus on differences across stars: we assume the observed dispersion in radii of protoplanetary disks reflect a range of initial conditions or different evolutionary pathways, and that those differences will be reflected in the observed semi-major axes of the observed exoplanet population.

To measure how the spatial scales of solids are distributed across stars in samples of both class II disks and exoplanets, we apply the same detection and selection bias corrections as in the preceding section in order to compare, as best as we can, detected solids around similar groups of stars.

\begin{figure}
    \centering
    \ifsubmit
    \includegraphics[width=\linewidth]{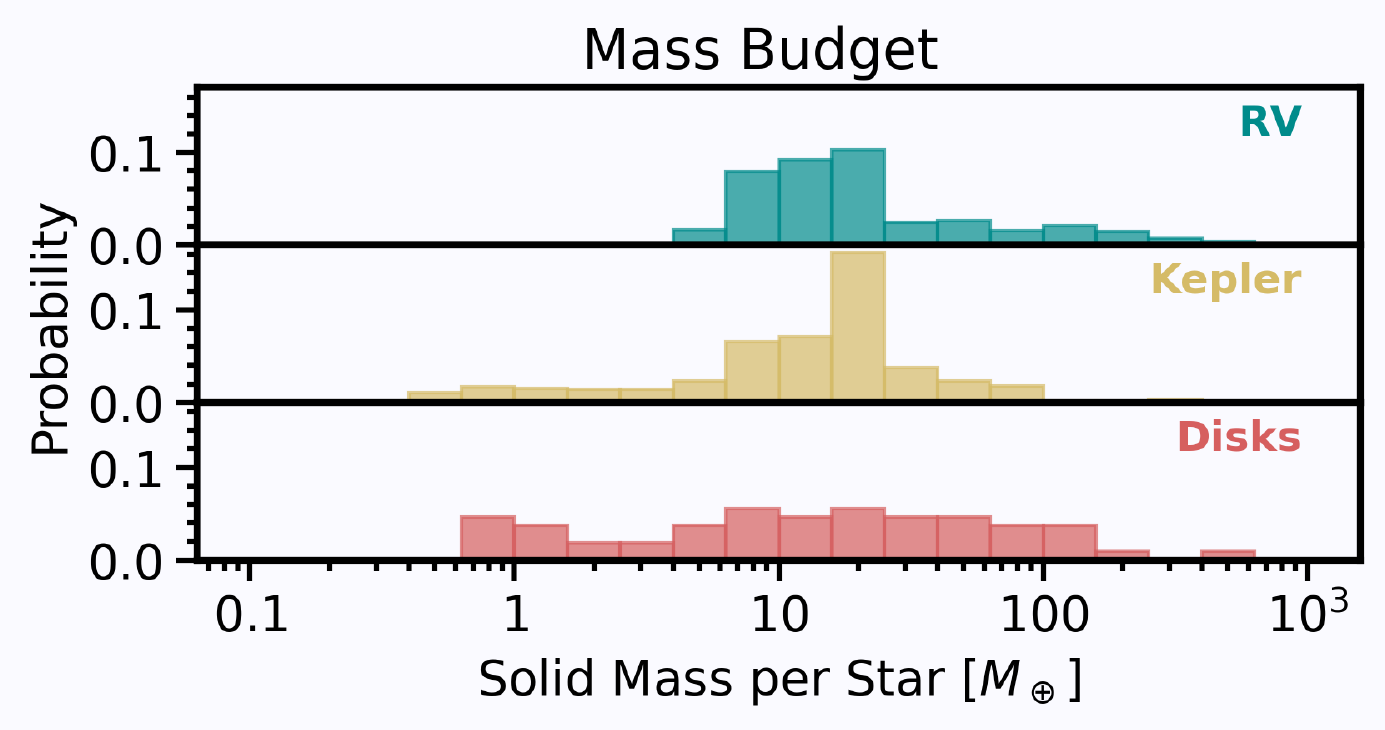}   
    \includegraphics[width=\linewidth]{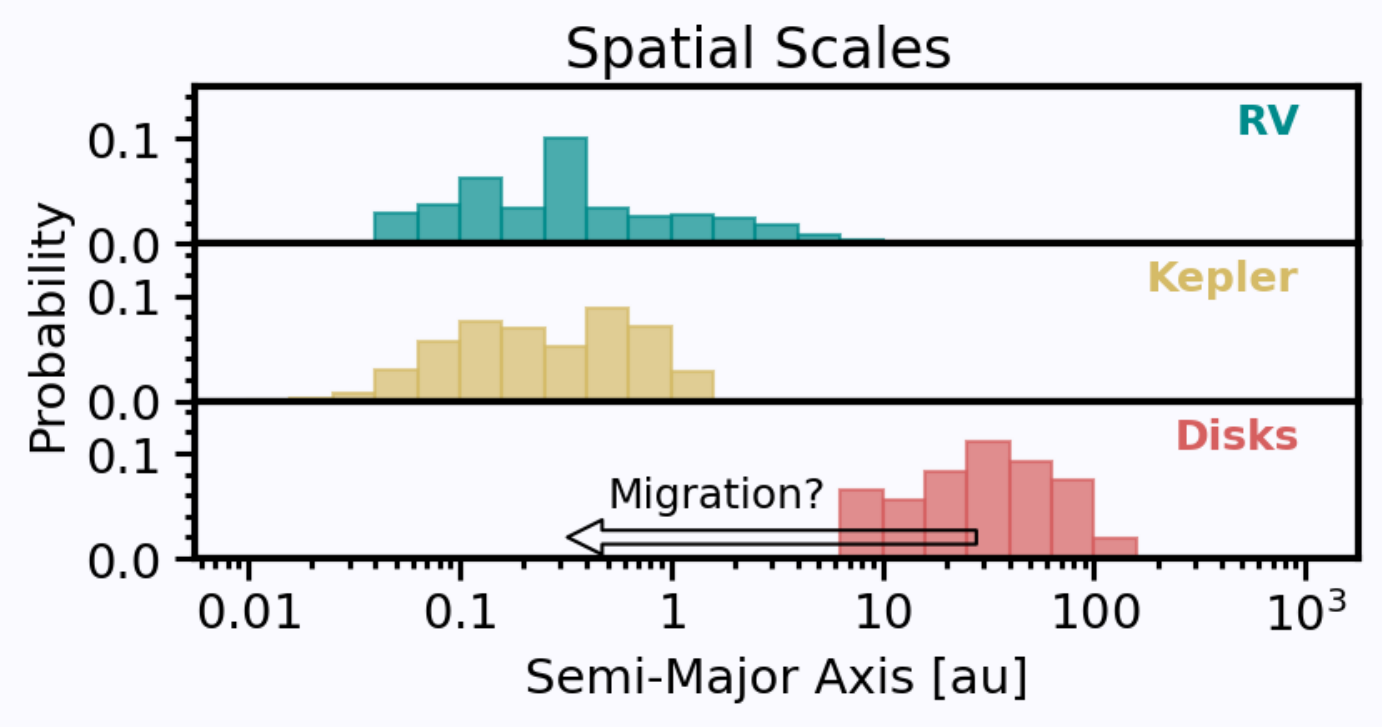}        
    \else
    \includegraphics[width=\linewidth]{fig_pdf/fig8a.pdf}   
    \includegraphics[width=\linewidth]{fig_pdf/fig8b.pdf}    
    \fi
    \caption{
    Completeness-corrected distribution of masses (a) and spatial scales (b) of solids in exoplanets and disks. 
    The arrow indicates the two orders of magnitude of planet migration or pebble drift needed if planets form exclusively from the solids currently observed in Class II disks. 
    }
    \label{f:hist}
\end{figure}

The results, described below, are visualized in Figures \ref{f:PR} and \ref{f:hist}. In these visuals and subsequent ones, the {occurrence of detected planets is explicitly accounted for, either by scaling the size of the symbols or by applying a weight in the histogram calculations.}

\subsection{Exoplanet Semi-Major Axis Distribution}
For exoplanets, the planetary system solid mass $M_s$ was estimated from each individual planet detection by accounting for the average number of planets per planetary system, $\bar{n}$. We continue following this approach here, and estimate the spatial scale where this mass is detected from the semi-major axis of that planet, which in turn is calculated from the planet orbital period using Kepler's third law and assuming a solar-mass star. 

These spatial scales are shown in Figure \ref{f:PR}, with each symbol representing a single detected planet.  The symbol area size for each planet is proportional to the weight factor, $w_j$, to illustrate the detection biases of these surveys. That is, large symbols correspond to planets detected despite a low survey detection efficiency, and thus those symbols represent a large number of undetected planets. Likewise, small symbols correspond to those planets that are easy to detect, and such planets are likely not underrepresented in our data.

The majority of planetary systems, those with solid masses in the range $5-20 M_\oplus$, are detected between 0.05 and 0.5 au. Their spatial distribution follows the known orbital period dependence of sub-Neptunes from the \kepler survey, where planet occurrence is measured to be roughly constant in the logarithm of orbital period between $10$ days and $1$ year, or $0.1-1$ au for a solar-mass star \citep[e.g.][]{2013PNAS..11019273P}. Inside of 0.1 au, the \kepler planet occurrence rate falls off rapidly with decreasing orbital period \citep[e.g.][]{2011ApJ...742...38Y,2012ApJS..201...15H}.
Planetary systems with solid masses above \mearth{20} are primarily detected in the radial velocity survey between $1-10$ au, consistent with estimates of \cite{2016ApJ...821...89B,2019ApJ...874...81F}. 
The onset of this rise in giant planet occurrence with semi-major axis \citep[e.g.][]{2013ApJ...778...53D,2016A&A...587A..64S} is also seen here in the \kepler data between $0.5-1$ au, near the edge of the detection limit at $\sim 1$ au.

Planetary systems with solid masses less the \mearth{5} are primarily detected within \au{0.2}. This is, at least in part, due to the detection bias of \kepler, whose minimum detectable planet size increases with orbital period, and is not sensitive to earth-sized planets outside of $\sim 50$ days. However, occurrence rate studies have also shown that smaller planets tend to be located closer in to the star than larger ones \citep{2012ApJS..201...15H,2018AJ....155...89P}, suggesting that the apparent correlation between spatial scale and planet size in Figure \ref{f:PR} may be a real feature of the exoplanet population {and not purely be attributed to survey biases and detection limits}. 

\begin{figure}
    \centering
    \ifsubmit
     \includegraphics[width=\linewidth]{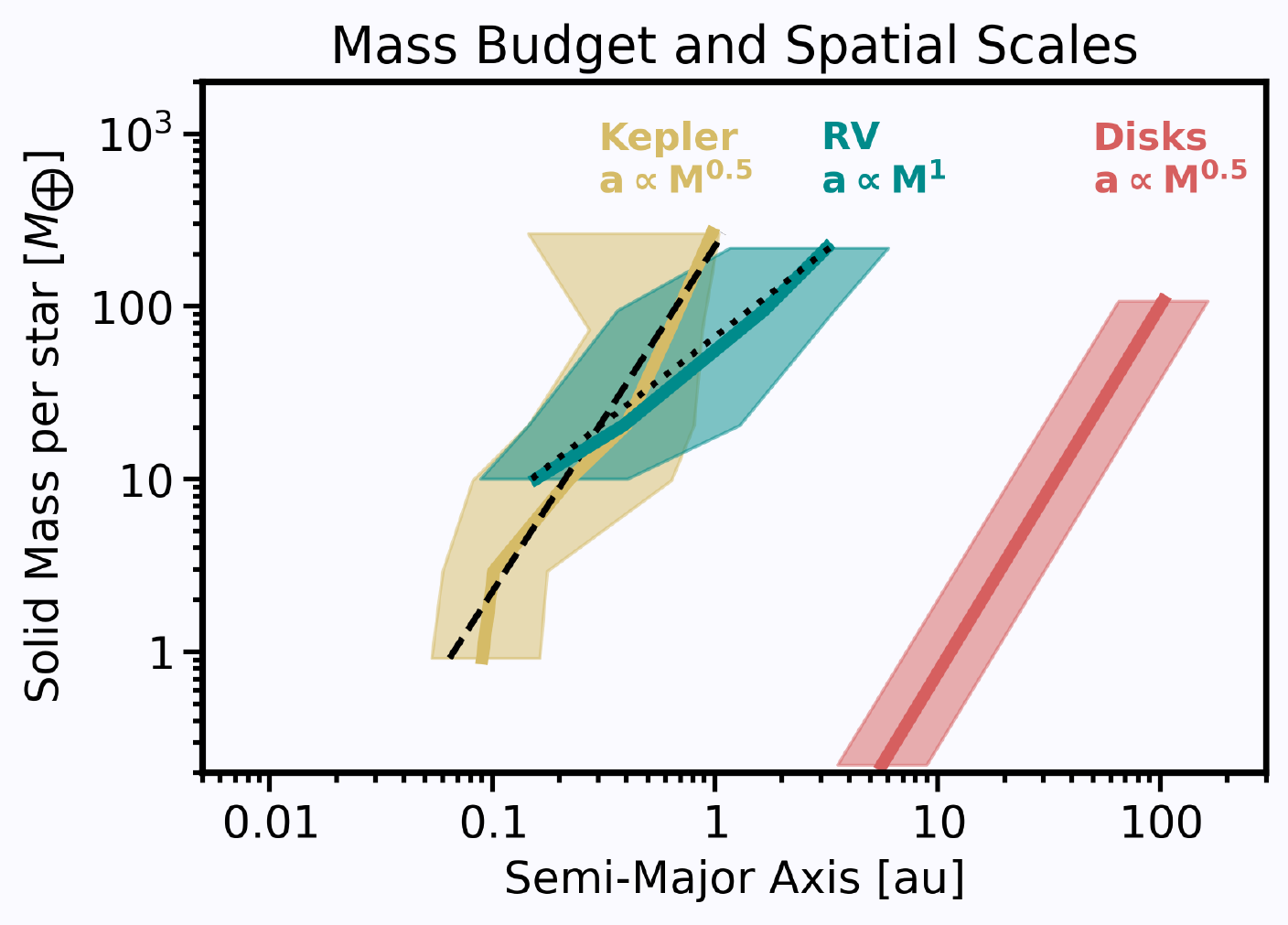}    
    \else
    \includegraphics[width=\linewidth]{fig_pdf/fig9.pdf}    
    \fi
    \caption{
    Quantile regression of the spatial scales of exoplanets systems at different system masses. The thick solid lines show the median while the shaded area shows the 1-$\sigma$ percentiles. The different slopes between the \kepler and RV data hints at detection limits playing a role in setting the exponent. However, the similarity with the disk size-mass relation is striking.
        }
    \label{f:quantiles}
\end{figure}

Overall, the spatial scales of detected exoplanets increase with planetary system solid mass, a trend also noted by other works \citep[e.g.][]{2019ApJ...875...29M}.
A quantile regression of the median planet semi-major axis at different {solid system masses, taking into account the detection efficiency weight factors $w$,} reveals that the \kepler planet population scales as $a \propto M_s^{0.5}$, while the radial velocity exoplanet population scales as $a \propto M_s^{1}$ \figp{quantiles}.
We note that the scaling derived here is significantly steeper than that of the Minimum Mass (Extra)Solar Nebula, which would be $M_s \propto a^{0...0.5}$ for surface density index p=1.5...2. This shows that this approach is indeed measuring variations in planetary system sizes among stars and not the surface density profiles of individual protoplanetary disks.

The detection limit of the combined \kepler and radial velocity exoplanet census roughly follows a steeper diagonal line that scales as $a \propto M_s^{1.35}$. As mentioned before, the exoplanet census for small ($<10\,M_\oplus$) planets is incomplete due to detection bias outside of $1$ au (and outside of $0.2$ au for earth-sized planets). While a decrease in the giant planet distribution with semi-major axis has been detected \citep{2019ApJ...874...81F,2019AJ....158...13N}, 
there is no clear indication from \kepler planet occurrence rate of a decline with semi-major axis near the detection limit at \au{1}, and thus the peak of this planet population could lie at larger semi-major axes. 
{Assuming an occurrence rate constant in the logarithm of orbital period, a similar amount of planets would be present between 1-10 au as detected with Kepler between 0.1-1 au.}
In that case, the apparent correlation between planet size and semi-major axis in Figures \ref{f:PR} and \ref{f:quantiles} could be the result of these detection limits. 
A scenario where the peak of planet occurrence of sub-Neptunes lies near that of the RV giants can neither be excluded nor confirmed with this data. 
Micro-lensing is sensitive to sub-Neptunes at separations between 1-10 au \citep{2012Natur.481..167C}, but their semi-major axis distribution has not been characterized in the context of a peaked distribution.
A census of small planets at larger semi-major axes, such as with the \textit{Nancy Grace Roman} Space Telescope 
is needed to confirm the relation between planet size and semi-major axis.

\subsection{Protoplanetary Disk Radii}
We calculate the spatial scale of the detected solid mass reservoir based on observed outer disk radii. Note that we do not consider the surface density profiles of individual disks, but that we use disk outer radii as a proxy for the spatial scale at which most solids are located.
The spatial distribution of dust in the outer regions of protoplanetary disks can be computed from the intermediate resolution ALMA data targeting all Class II disks in Lupus and Chameleon I \citep{2016ApJ...828...46A,2016ApJ...831..125P,2018ApJ...863...61L,2018ApJ...859...21A}

Here, we use dust disk outer radii from the analysis of \cite{2020ApJ...895..126H}, who fitted radial profile models to visibility data of a large sample of Class II disks with available ALMA interferometric data, and reported the radii that enclose 68\% and 90\% of the millimeter flux. We take the radii that encloses 68\% to be a tracer of where most of the disk mass is located.
The flux distribution is a proxy for the solid mass distribution, with the two related by the dust temperature and optical depth. The location that encloses 68\% of the flux encloses less than 68\% of the mass because temperature increases inward. 
Therefore, we take the $R_{68}$ to be {a proxy} of $M_{50}$, e.g the point where half of the mass is detected. A full radiative transfer calculation to estimate the disk surface density is beyond the scope of this paper. However, $R_{68}$ and $R_{90}$ are tightly correlated and their offset is only 0.1 dex \citep{2020ApJ...895..126H}, and thus the choice for one over the other would not significantly affect the conclusions of this paper. 

From the sample of 54 Sun-like stars ($0.5 < M/M_\odot < 2$) in Lupus and Chamaeleon I used in Section \ref{s:solids}, 36 have a disk radius measurement of which 6 are an upper limit. 
We calculate the disk mass from the millimeter flux using the same disk temperature of $T=20\, K$ as in \cite{2016ApJ...831..125P}. To verify that the sample of disks with radius measurements is not biased towards bigger, brighter disks compared to the entire sample that also includes disks without radius measurements, we calculate the KS distance between the two mass distributions and find a high ($p_\text{KS} = 83\%$) probability that the two samples are drawn from the same distribution. 

The disk masses and radii are shown in red in Figure \ref{f:PR}.
Upper limits are indicated with triangles. 
The disk mass and outer radii are positively correlated \figp{quantiles}, with $R_{68} \propto M_d^{0.5}$ and a scatter of 0.2 dex \citep[see also][]{2017ApJ...845...44T,2018ApJ...865..157A}.

\subsection{Comparing Disk and Exoplanet Spatial Scale}
The comparison between spatial scales of detected solids in protoplanetary disks and exoplanets does not significantly change the need for inward migration of pebbles or proto-planets:
The spatial scales where solids in protoplanetary disks are detected are two orders of magnitude larger than those of exoplanets. Most of the solids in exoplanet systems are detected between 0.1-1 au, while most of the disk emission {that represents a similar amount of solids} is detected between 10-100 au. This is indicated with an arrow in Figure \ref{f:hist}b. If observed exoplanets indeed form from the observed mass reservoir in protoplanetary disks, then planets or their building blocks need to migrate inward over two orders of magnitude during formation. Both planet migration and a pebble accretion scenario could achieve this.

The comparison does show a hint of how this migration mechanism operates:
detected exoplanets and disks both show a positive correlation between the amount of solids and the semi-major axis where they are detected \figp{quantiles}. 
More massive planets are located farther form the star, and the disks that contain enough solids to form them are also larger. 
This suggests that the observed size scaling of the solid mass reservoir may be retained during the planet formation process, even if exoplanets or their building blocks migrate inward by a factor 100 in semi-major axis.

Within the hypothesis {that planets migrate inward through type I or type II migration}, it is not clear why such a relation would be preserved, as the final locations of exoplanets are determined more by the location where migrating planets are trapped, and less by their starting location.
Within the pebble accretion hypothesis, a similar argument can be made that the fast radial drift of pebbles could erase this scaling relation, and that the locations of exoplanets are determined more by where pebbles are accreted, and not where they came from. However, since the observed Class II disk dust distributions may already be sculpted by drift of pebbles \citep[e.g.][]{2020A&A...635A.105P,2021AJ....162...28V}, this could be seen as an argument in favor of pebble accretion.

\section{Summary and Discussion}\label{s:discuss}
We have calculated the mass and spatial distribution of heavy elements contained in exoplanets around Sun-like stars from the \kepler and radial velocity surveys,
and compared those to the solids detected in Class II protoplanetary disks in the Lupus and Chamaeleon I star forming regions. 
The solid mass reservoirs of exoplanets and protoplanetary disks appear to be of similar magnitude, consistent with previous results \citep{2014MNRAS.445.3315N}. 
We find that roughly half of Sun-like stars have disks and exoplanets, with a median solid mass of $\approx 10-20 M_\oplus$.

The solid mass reservoirs trace different spatial scales, consistent with the idea that planets or their building blocks migrate inward during formation.
Most of the solids contained in exoplanets are located between 0.1 and 1 au, with a smaller contribution traced by giant planets between 1-10 au. 
In contrast, the same amount of dust in protoplanetary disks is detected at spatial scales of 10-100 au, e.g. two order of magnitude larger.

The correlation between protoplanetary disk radius and mass \citep[e.g.][]{2017ApJ...845...44T} is possibly reflected in the exoplanet population, with the more massive giant planets preferentially located farther out than sub-Neptunes. However, a census of sub-Neptunes between 1 and 10 au from the star, such as with the Nancy Grace Roman Space Telescope \citep[e.g.][]{2019ApJS..241....3P},
would be needed to confirm this is not due to detection bias. If confirmed, the locations where exoplanets form may have a direct dependence on the initial distribution of (solid) mass in the disk, providing a direct link between the properties of protoplanetary disks and the variety of observed exoplanet systems. 

The results of the analysis presented here are in broad lines consistent with that of previous papers. Protoplanetary disks around Sun-like stars do contain similar amounts of solid material as contained in exoplanets \citep{2014MNRAS.445.3315N,2015ApJ...814..130M,2016ApJ...831..125P}. 
We do not find support for the statement in \cite{2018A&A...618L...3M} that ``exoplanetary systems masses are comparable or higher than the most massive disks''.
Instead, a discrepancy between disks and exoplanets arises only when considering either an additional population of planets not yet detected around Sun-like stars or when considering that the efficiency with which protoplanetary disks convert their solids into exoplanets is likely less than 100\%. 

\begin{figure}
    \centering
    \ifsubmit
    \includegraphics[width=\linewidth]{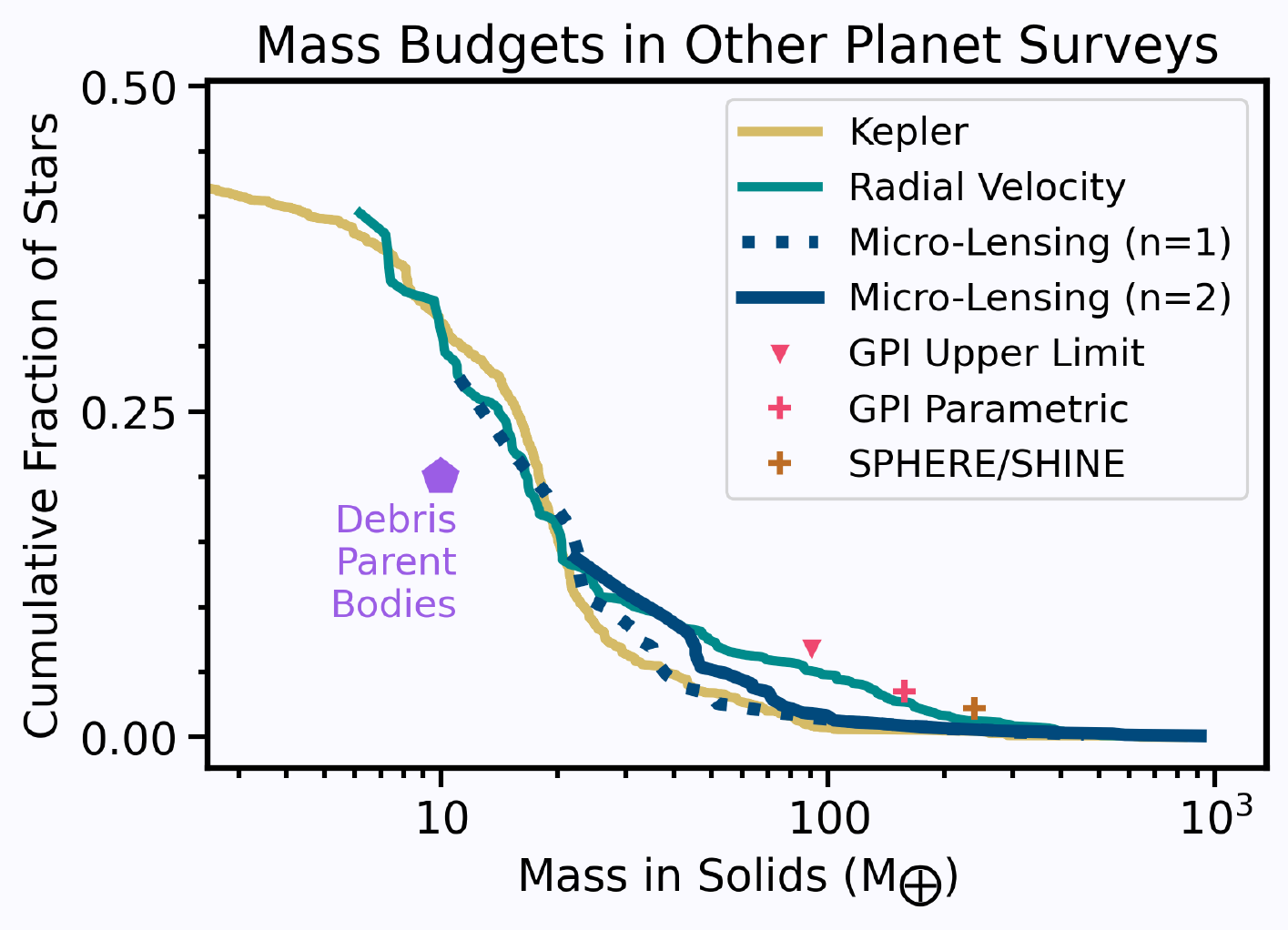}
    \else
    \includegraphics[width=\linewidth]{fig_pdf/fig10.pdf} 
    \fi
    \caption{
	Solid mass reservoirs around sun-like stars based on other exoplanet surveys compared to the distributions from Fig. \ref{f:cdf}.
	The mass in Microlensing planets is based on \citep{2016ApJ...833..145S} assuming solar-mass host stars and for two different assumptions of the number of planets per system ($n$) in blue. 
	The mass in directly imaged planets is based on the single exoplanet detection around FGK stars from SPHERE/SHINE (Brown Cross, \citealt{2021A&A...651A..72V}) and the upper limits and parametric models from GPI (Pink, \citealt{2019AJ....158...13N}). 
	The mass distribution of parent bodies of debris disks is not directly measured, but an estimate of $10 M_\earth$ \citep{2021MNRAS.500..718K} for $20\%$ of G stars \citep{2013A&A...555A..11E} is indicated with a purple pentagon.  
	}
    \label{f:otherplanets}
\end{figure}

\begin{figure}
    \centering
    \ifsubmit
     \includegraphics[width=\linewidth]{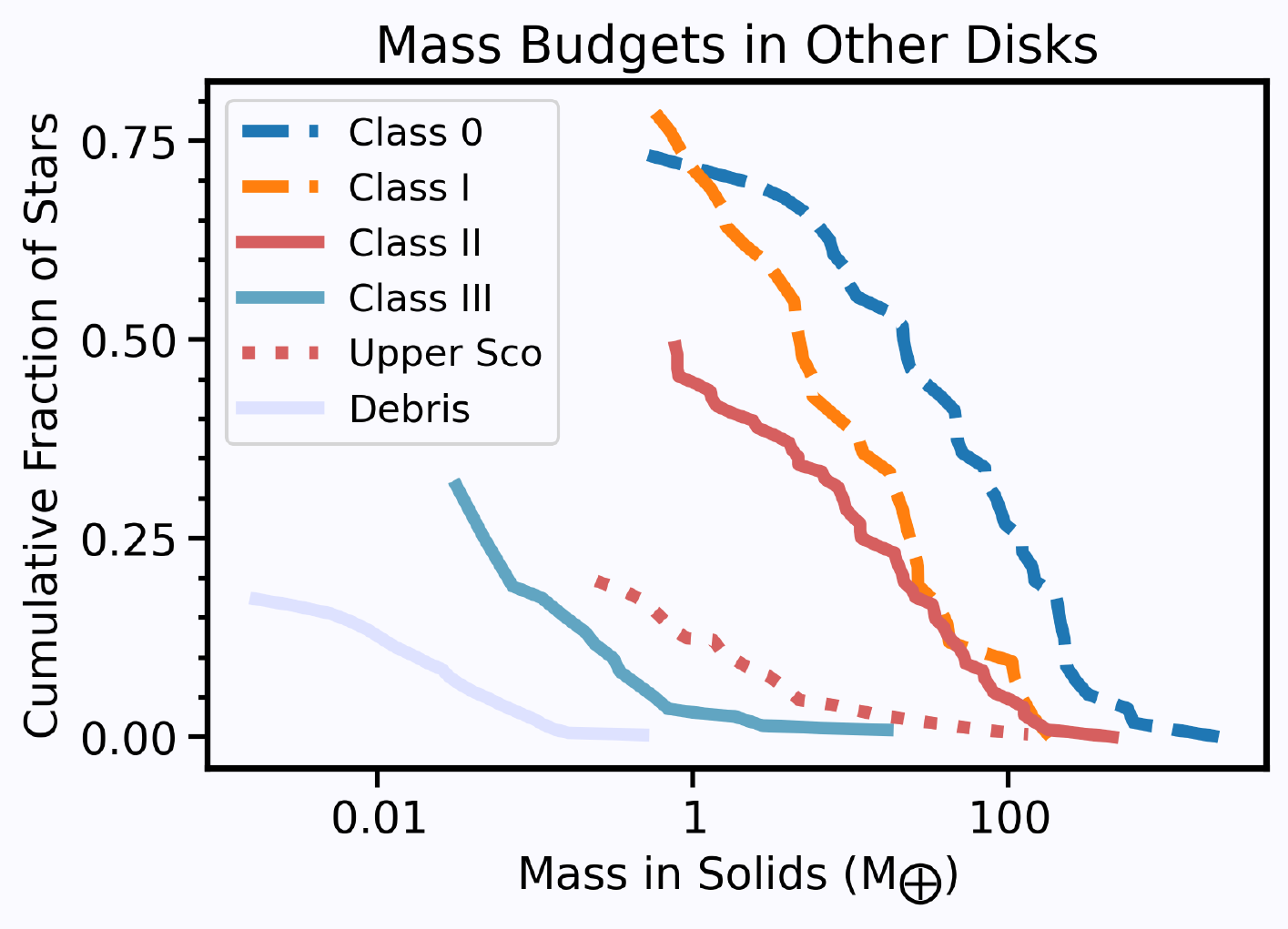} 
    \else
    \includegraphics[width=\linewidth]{fig_pdf/fig11.pdf} 
    \fi
    \caption{
	Solid mass reservoirs from other disk surveys compared to the distributions from Fig. \ref{f:cdf}.
	The masses of younger disks, Class 0 and I, are from \cite{2020A&A...640A..19T} and have no measured stellar masses.
	The solid mass distributions of older disks are based on the compilation of \cite{2021arXiv210405894M}, and adjusted for the disk fraction and scaled to solar mass stars if needed.
	Debris disks masses are from \citep{2017MNRAS.470.3606H} scaled to a 20\% disk fraction. 
	Class III disks masses are from \cite{2021arXiv210405894M} scaled to 50\% of stars.
	The masses of older Class II disks in Upper Sco are from \citep{2016ApJ...827..142B} and scaled to a disk fraction of 25\%.
	}
    \label{f:otherdisks}
\end{figure}

There are three datasets that provide additional constraints on the planet population. We have not included them in this analysis because they do not directly measure the solid mass distribution around sun-like stars. However, we show how these results complement our analysis in Figures \ref{f:otherplanets} and \ref{f:otherdisks}:
\begin{enumerate}
\item Micro-lensing surveys have detected a significant population of Neptune-mass planets at a few au around predominantly low-mass stars \citep[e.g.][]{2012Natur.481..167C}. The planets were mainly thought to orbit low-mass M dwarfs, and because the host star masses have not been directly measured in many cases, it is not clear whether this sample also represents a direct measurement of the planet occurrence rates around solar-mass stars.
In figure \ref{f:otherplanets} we show the microlensing planetary system mass distribution using the detection efficiency and planet/star mass ratio from \cite{2016ApJ...833..145S}, assuming solar-mass host stars and two planets per planetary system. These rates are consistent with the radial velocity solid mass distribution.
\item Roughly 20\% of sun-like stars have detected debris disks \citep{2009ApJ...705.1646C,2013A&A...555A..11E,2018MNRAS.475.3046S}.
The detected mass in debris disks (and also class III disks) is typically smaller than that in class II disks \citep[e.g.][]{2017MNRAS.470.3606H}, see also Figure \ref{f:otherdisks}. However, this debris likely corresponds to a large mass reservoir of solids outside of 20 au needed to continually replenish the detected small dust. 
\cite{2021MNRAS.500..718K} estimate using models that the mass reservoir of parent bodies is of order $\sim 10 M_\oplus$ (see also \citealt{2010MNRAS.405.1253K, 2010ApJS..188..242K}). This is indicated with a purple pentagon in figure \ref{f:otherplanets}. It is not yet clear whether these parent bodies reside in the same systems as detected exoplanets \citep{2020MNRAS.495.1943Y} or whether they form a separate reservoir of material that would increase the fraction of stars with solids to $\approx 70\%$. Currently, there is no sample of directly estimated planetesimal belt masses (and semi-major axes) for a representative range of debris disks for sun-like stars to include in our analysis.

\item Direct imaging surveys have provided constraints on the occurrence rate of super-Jupiters at large separations \citep[e.g.][]{2019AJ....158...13N,2021A&A...651A..72V}. The SPHERE/SHINE survey has detected one 10 Jupiter mass exoplanet around an FGK star, corresponding to an occurrence rate of 2.5\% \citep{2021A&A...651A..72V}.
We show this detection in Figure \ref{f:otherplanets}, which we assign a solid mass of $M_s= 238 M_\oplus$ following Eq \ref{eq:mcore}. The GPI survey has reported an upper limit of 6.9\% for exoplanets around FGK stars with masses larger than 2 Jupiter mass \citep{2019AJ....158...13N}. 
Most directly imaged planets ($<13 M_J$) are detected around intermediate-mass stars ($>1.5 M_\odot$), but conclusions about the planet population around sun-like stars can also be drawn by fitting distributions in planet/brown dwarf mass, semi-major axis, and stellar mass. These distributions are generally consistent with  radial velocity mass and semi-major axis distribution extrapolated to larger radii \citep[e.g.][]{2019ApJ...874...81F,2019AJ....158...13N}.
We also show the estimated 3.5\% occurrence of the parametric model from GPI for planets larger than 5 Jupiter masses ($M_s= 158$) in figure \ref{f:otherplanets}. 
All these limits are consistent with the tail of the radial velocity planet mass distribution.
\end{enumerate}

Inward migration of planets or planetary building blocks may seem a likely explanation to the discrepancy in spatial scales, as those mechanisms have the potential to move significant amount of material inward over large distances at short time scales. However, doing so would require those mechanisms to convert nearly all of the observed dust into exoplanets. In reality, those mechanisms are likely to operate at a much lower efficiency, as also pointed out by \cite{2014MNRAS.445.3315N,2018A&A...618L...3M}. 
Planet population synthesis models that match observed exoplanet population using disk-migration \citep{2018haex.bookE.143M,2019ApJ...887..157M} start with a population of disks that are significantly more massive, because most planetesimals are not accreted onto planets. 
Pebble accretion models also require massive disks because most pebbles are not accreted onto planets but drift into the star \citep[e.g.][]{2020A&A...638A.156A}.
Thus, while the observed solid material in Class II protoplanetary disks is large enough to match current-day exoplanets, the reservoir from which exoplanets formed was likely larger.
Surveys of younger, more embedded Class 0/I disks have indeed identified that these disks appear to be significantly more massive \citep{2018ApJS..238...19T,2020ApJ...890..130T,2020A&A...640A..19T}. 
We show the derived dust mass distributions of Class 0 and Class I objects from \cite{2020A&A...640A..19T} in figure \ref{f:otherdisks}. The stellar mass distribution and disks fraction of these objects are not well determined, and thus we can not correct for the selection bias in this sample.

A big caveat to our analysis of the solid mass reservoirs would be the case that the masses of Class II disks are systematically under-estimated. The conversion from millimeter flux to dust mass rests on assumptions about the dust properties, disk temperature, and optical depth. 
Some of these uncertainties, in particular on disk temperature and optical depth, can be mitigated using radiative transfer modeling \citep[e.g.][]{2019AJ....157..144B}. However, the opacity of dust, reflecting both its composition and grain size distribution, remain uncertain to a large degree \citep[e.g.][]{2018ApJ...869L..45B}.
Recently, \cite{2019ApJ...877L..18Z} found indications that the dust emission traced with ALMA at millimeter wavelengths is optically thick, and thus their mass and surface density could be significantly underestimated, as also noted in \cite{2018ApJ...865..157A}. If disks have an additional mass reservoir in the inner disk that planets form from, this would likely go undetected in ALMA surveys. 
 
Finally, we have focussed the comparisons in this paper on Sun-like stars, mainly because this is where most exoplanets to date have been detected, and population statistics are available for a wide range of planet masses and orbital separations, and from different detection methods. Most of the protoplanetary disks in nearby star-forming regions, however, orbit lower mass stars that will become main sequence M dwarfs. The census of exoplanets around these lower-mass stars is less complete due to these stars being intrinsically less bright and therefore being under-represented in magnitude limited surveys \citep[e.g.][]{1922MeLuF.100....1M}. However, a number of key differences in the exoplanet population compared to that around Sun-like stars have emerged from smaller surveys.
M dwarfs have a higher occurrence rate of planets smaller than Neptune \citep[e.g.][]{2015ApJ...798..112M}, but a lower occurrence rate of giant planets \citep[e.g.][]{2010PASP..122..905J}. Micro-lensing surveys have also revealed a large number of Neptunes-mass planets near the snow line around low-mass stars \citep[e.g.][]{2016ApJ...833..145S}. 
These observations appear to challenge planet formation models more than exoplanets around Sun-like stars because their protoplanetary disks are expected to be lower in mass \citep[e.g.][]{2015ApJ...814..130M,2017MNRAS.470L...1G,2018ApJ...869L..34S}.
While it is beyond the scope of this paper to create a planet population model for M dwarfs, it would be worth reevaluating these exoplanet population properties in the light of the new correlations between disk size, disk mass, and stellar mass emerging from ALMA surveys \citep[e.g.][]{2018ApJ...865..157A,2020ApJ...895..126H}. 
In a separate paper, \cite{2021AJ....162...28V}, we propose a hypothesis for the planet population around M dwarfs based on the statistics of gaps and rings detected in protoplanetary disks with ALMA.

\begin{acknowledgments} 
This paper paper benefitted from discussions within the ``Zooming In On Rocky Planet Formation" team at the International Space Science Institute in Bern, Switzerland. 
We thank Nienke van der Marel and {\L}ukasz Tychoniec for providing dust disk masses in electronic format.
G.D.M. would like to thank the members of the Ciesla group for feedback and moral support during the writing process, and for their relentless criticism of the color schemes used in this paper.

G.D.M. acknowledges support from ANID --- Millennium Science Initiative ---  ICN12\_009.
This material is based upon work supported by the National Aeronautics and Space Administration under Agreement No. 80NSSC21K0593 for the program ``Alien Earths''. The results reported herein benefitted from collaborations and/or information exchange within NASA's Nexus for Exoplanet System Science (NExSS) research coordination network sponsored by NASA’s Science Mission Directorate.\\
\end{acknowledgments}

\software{
NumPy \citep{numpy}
Matplotlib \citep{pyplot}
SciPy \citep{scipy}
Astropy \citep{astropy}
\texttt{epos} \citep{epos}
KeplerPORTs \citep{2017ksci.rept...19B}
}

\ifsubmit
	\bibliography{scales}
\else	
	\bibliography{papers3,books,software,inprep}
\fi


\end{document}